\documentclass[12pt]{iopart}

\usepackage{iopams}
\usepackage{isomath}
\usepackage{dsfont}
\usepackage{graphicx}
\usepackage{bm}
\usepackage{amssymb}
\usepackage{mathrsfs}
\usepackage{color}
\usepackage{cancel}
\usepackage{tikz}
\usepackage{tkz-euclide}
\usepackage{tikz-3dplot}
\tdplotsetmaincoords{70}{110}
\usepackage{tabularx}
\usepackage{pgfplots}
\usepgfplotslibrary{groupplots}
\pgfplotsset{compat=1.7}
\usepgfplotslibrary{patchplots}
\usepgfplotslibrary{fillbetween}
\usepackage{bibentry}
\usetikzlibrary{decorations}
\usetikzlibrary{math} 
\usetikzlibrary{decorations.markings}
\usepackage{caption} 
\usepackage{float} 
\usepackage{multirow} 
\usepackage{subcaption} 
\usepackage{subcaption}

\usepackage{xcolor}
\definecolor{OffWhite}{HTML}{FFF7DD}
\definecolor{DarkGreen}{HTML}{016401}
\definecolor{DarkRed}{HTML}{B30900}
\definecolor{DarkBlue}{HTML}{0047AB}
\definecolor{BrightBlue}{HTML}{04D9FF}
\pgfplotsset{
	colormap={slategraywhite}{
		rgb255=(112,112,112)
		rgb255=(255,255,255)
	},
}
\pgfmathdeclarefunction{gaussian}{2}{%
	\pgfmathparse{exp(-((x-#1)^2)/(2*#2^2))}%
}

\begin{document}

\title[Conceptual study on using Doppler backscattering to measure magnetic pitch angle in tokamak plasmas]{Conceptual study on using Doppler backscattering to measure magnetic pitch angle in tokamak plasmas}

\author{AK Yeoh$^{1, 2}$, VH Hall-Chen$^{1}$, QT Pratt$^{3}$, BS Victor$^{4}$, J Damba$^{3}$, TL Rhodes$^{3}$, NA Crocker$^{3}$, KR Fong$^{1, 5}$, JC Hillesheim$^{6}$, FI Parra$^{7}$ and J Ruiz Ruiz$^{8}$}

\address{
	$^1$Future Energy Acceleration and Translation Programme, Agency of Science, Technology and Research (A*STAR), Singapore 138632, Singapore\\
	$^2$Department of Physics, University of Oxford, Oxford OX1 3PU, UK\\
	$^3$Department of Physics and Astronomy, University of California, Los Angeles, CA 90095, USA \\
	$^4$Lawrence Livermore National Laboratory, Livermore, CA 94551, United States\\
	$^5$Department of Physics, National University of Singapore, Singapore 117551, Singapore\\
	$^6$Commonwealth Fusion Systems, Cambridge, MA, USA \\
	$^7$Princeton Plasma Physics Laboratory, Princeton, NJ 08540, USA\\
	$^8$Rudolf Peierls Centre for Theoretical Physics, University of Oxford, Oxford, OX1 3NP, UK\\}
\ead{valerian\_hall-chen@a-star.edu.sg}
\vspace{10pt}
\begin{indented}
\item[]November 2025
\end{indented}

\begin{abstract}
	We introduce a new approach to measure the magnetic pitch angle profile in tokamak plasmas with Doppler backscattering (DBS), a technique traditionally used for measuring flows and density fluctuations. The DBS signal is maximised when its probe beam's wavevector is perpendicular to the magnetic field at the cutoff location, independent of the density fluctuations [Hillesheim \emph{et al} 2015 \emph{Nucl. Fusion} \textbf{55} 073024]. Hence, if one could isolate this effect, DBS would then yield information about the magnetic pitch angle. By varying the toroidal launch angle, the DBS beam reaches cutoff with different angles with respect to the magnetic field, but with other properties remaining similar. Hence, the toroidal launch angle which gives maximum backscattered power is thus that which is matched to the pitch angle at the cutoff location, enabling inference of the magnetic pitch angle. We performed systematic scans of the DBS toroidal launch angle for repeated DIII-D tokamak discharges. Experimental DBS data from this scan were analysed and combined with Gaussian beam-tracing simulations using the Scotty code [Hall-Chen \emph{et al} 2022 \emph{Plasma Phys. Control. Fusion} \textbf{64} 095002]. The pitch-angle inferred from DBS is consistent with that from magnetics-only and motional-Stark-effect-constrained (MSE) equilibrium reconstruction in the edge. In the core, the pitch angles from DBS and magnetics-only reconstructions differ by one to two degrees, while simultaneous MSE measurements were not available. The uncertainty in these measurements was under a degree; we show that this uncertainty is primarily due to the error in toroidal steering, the number of toroidally separated measurements, and shot-to-shot repeatability. We find that the error of pitch-angle measurements can be reduced by optimising the poloidal launch angle and initial beam properties. Since DBS has high spatial and temporal resolutions, is non-perturbative, does not require neutral beams, and is likely robust to neutron damage of and debris on the first mirrors, using DBS to measure the pitch angle in future fusion energy systems is especially appealing.
\end{abstract}
%
\vspace{2pc}
\vspace{2pc}
\newline
%
\submitto{\NF}
%
%
%
\section{Introduction} 
\label{sec:Introduction}
Measuring magnetic fields in the core of fusion plasmas is important for determining magnetohydrodynamic (MHD) stability and predicting disruptions \cite{Sweeney:MHD:2020}, calculating turbulent transport \cite{Burrell:shear:1997}, understanding Alfv{\'e}n eigenmodes, understanding fast ion modes and losses \cite{Cheng:alfven:1985, Fasoli:MHD:2002, Gorelenkov:alfven:2024}, and validating models of current drive \cite{Mumgaard:LHCD:2015}. These measurements will play a vital role on next-generation tokamaks like ITER \cite{Foley:MSE:2008}, SPARC \cite{Stewart:SPARC_magnetics:2023}, BEST, CFETR \cite{Hu:CFETR:2020}, and STEP \cite{Anand:STEP:2023} operating in the burning plasma regime and predicting performance in future fusion energy systems. Ideally, magnetic field measurements would: 
\begin{itemize}
	\item Be direct measurement of the magnetic field instead of time-integrated $\rmd \mathbf{B} / \rmd t$ measurements, as the latter can be susceptible to drifts over longer pulse durations;
	\item Have high spatial resolution and access to local measurements in the plasma core;
	\item Have high time resolution;
	\item Have measurement hardware robust to sustained operations under high temperature and neutron loading conditions;
	\item Be operable without relying on other systems, such as neutral beams.
	\item Be non-perturbative.
\end{itemize}
Existing techniques of measuring the pitch angle, such as magnetic coils \cite{Raju:mirnov:2000, Hole:mirnov:2009}, motional-Stark effect (MSE) \cite{Levinton:MSE:1990, Wroblewski:MSE:1992, Levinton:MSE:1999, Gibson:MSE:2021, Conway:MSE:2010, Wolf:MSE:2015}, and polarimetry \cite{Chen:polarimetry:2022}, are unable to fully meet the requirements of a magnetic field diagnostic for a future fusion energy system as set forth here. Magnetic coils measure $\rmd \mathbf{B} / \rmd t$ and are expected to suffer from neutron damage. MSE requires neutral beams, which are highly perturbative, and as an optical technique MSE is sensitive to debris on the first mirror. Finally, while polarimetry has many advantages, it can only provide line-averaged measurements.

In this paper, we show how the Doppler backscattering (DBS) diagnostic can be used to measure the magnetic pitch angle, potentially satisfying the list of requirements in the preceding paragraph,
with the exception of requiring electron density measurements. Nonetheless, since profile reflectometers to measure electron density are expected to be a mainstay of future devices \cite{Vayakis:ITER:2006, Orsitto:DEMO:2016, Lin:SPARC:2024}, no additional dedicated supporting systems are needed. 

While DBS is traditionally used to measure flows and turbulent density fluctuations \cite{Hennequin:DBS:2004, Happel:DBS:2010, Shi:DBS:2016, Rhodes:DBS:2016, Hu:DBS:2017, Tokuzawa:DBS_LHD:2021, Estrada:DBS:2021, Ren:DBS:2021, Yashin:DBS:2022, Pratt:DBS:2022, Rhodes:DBS:2022, Shi:DBS:2023, Pratt:spectrum:2023, Chowdhury:DBS:2023}, recent work \cite{Hall-Chen:beam_model_DBS:2022, Hall-Chen:mismatch:2022, Damba:mismatch:2022, Hall-Chen:mismatch:2024} indicates that the variation of backscattered signal, from simultaneous toroidally separated measurements at the same radial location, might enable one to infer the magnetic pitch angle. We further develop this approach in this paper. Here, the magnetic pitch angle is given by $\gamma = \tan^{-1} \left(B_{\rm p} / B_\zeta \right) $, where $B_{\rm p}$ is the poloidal field and $B_\zeta$ is the toroidal field. This magnetic pitch angle is related to the safety factor, which is important in the theory of MHD stability and Alfv{\'e}n eigenmodes.

DBS measurements typically rely on complicated analysis techniques to measure the Doppler shift, which yields information about the background flow \cite{Happel:DBS:2010, Pratt:DBS:2022}. However, our proposed technique only relies on the backscattered power, which can be calculated without having to find the Doppler shift. This simplified analysis makes our approach more amenable to implementation of this diagnostic for real time control in future machines. Moreover, as a microwave diagnostic, DBS is robust against neutron damage \cite{Orsitto:DEMO:2016, Volpe:microwave:2017}. By using microwave waveguides, the DBS electronics can be placed behind neutron shielding. Moreover, microwave techniques' longer wavelengths (compared to optical) make them more tolerant to larger defects, such as dust on first mirrors. While there have been other microwave-based techniques for measuring the magnetic field \cite{Gourdain:assessment:2008, Gourdain:application:2008, Thomas:SAMI:2016, Prisiazhniuk:PCDR:2017, Allen:SAMI2:2021}, our approach requires either simpler hardware or simpler analysis techniques. 

DBS involves sending a microwave probe beam into the plasma. This beam is typically strongly refracted, which is believed to localise the measured backscattered signal to the turning point, which we refer to as the nominal cutoff position. 
By fixing the poloidal launch angle and varying the toroidal launch angle, for a given DBS frequency channel, we expect to reach a similar cutoff location with a similar wavenumber at cutoff. The initial wavevector, $\mathbf{K}_{\rm ant}$, depends on the launch angles,
\begin{equation} \label{eq:launchK}
	\eqalign{
		K_{\rm R, ant} &= - \frac{\Omega}{c} \cos \varphi_{\rm t} \cos \varphi_{\rm p} , \\
		K_{ \zeta \rm, ant} &= - \frac{\Omega}{c} R \sin \varphi_{\rm t} \cos \varphi_{\rm p} , \\
		K_{\rm Z, ant}  &= - \frac{\Omega}{c} \sin \varphi_{\rm p} .
	}
\end{equation}
Here $\Omega$ is the angular frequency of the probe beam, $c$ is the speed of light, position is given in the right-handed $(R, \zeta, Z)$ cylindrical basis, and $\varphi_{\rm p}$ and $\varphi_{\rm t}$ are the poloidal and toroidal launch angles, respectively. The angle between the probe beam's wavevector and the plane perpendicular to the magnetic field is called the mismatch angle, given by
\begin{equation}
\label{eq:MismatchAngle}
	\sin \theta_{\rm m} = \frac{K_\parallel}{K},
\end{equation}
where $\mathbf{K}$ is the wavevector of the probe beam, $K = |\mathbf{K}|$, $K_\parallel = \mathbf{K} \cdot \hat{\mathbf{b}}$, and $\hat{\mathbf{b}}$ is the unit vector of the equilibrium magnetic field. While there is a mismatch angle associated with every point along the path of the beam, we focus on the mismatch angle at the cutoff in this paper. The backscattered signal is strongly attenuated by this mismatch, which in turn depends on the toroidal launch angle, as shown in previous experimental and theoretical work \cite{Hillesheim:DBS_MAST:2015, Hall-Chen:beam_model_DBS:2022, Hall-Chen:mismatch:2022, Damba:mismatch:2022, Hall-Chen:mismatch:2024}. From here on, we refer to the variation in received backscattered power with toroidal launch angle as the toroidal response. Subsequently, we refer to the toroidal launch angle which maximises the backscattered power as the optimal toroidal angle, $\varphi_{\rm t,opt}$. For the shots studied, the backscattered power is maximised when $\theta_{\rm m} = 0$, see \ref{sec:wavenumberSpectrum}. Since the mismatch angle is related to orientation of the magnetic field; one should be able to infer the pitch angle at the cutoff location, see Figure \ref{fig:ToroidalSteering}. In broad strokes, this is the principle of using DBS to measure the magnetic pitch angle. Using DBS data from DIII-D \cite{Fenstermacher:DIIID:2022, Holcomb:DIIID:2024}, we show that inferring the magnetic pitch angle with this approach is indeed possible.

\begin{figure*}
	
	\centering
	
	\scalebox{1.0}{
		\begin{tikzpicture}
			\begin{axis}[
				axis equal image,
				hide axis,
				z buffer = sort,
				view = {0}{20},
				scale = 1.5, 
				xmin = -5.0,
				xmax = 4.1,
				ymin = -4.1,
				ymax = 3.1,
				zmin = -3.1,
				zmax = 3.1,
				]
				\addplot3[
				surf,
				shader = faceted interp,
				samples = 17,
				samples y = 33,
				domain = -2*pi:0,
				domain y = -2*pi:0,
				colormap name = slategraywhite,
				thin,
				](
				{(3+sin(deg(\x)))*cos(deg(\y))},
				{(3+sin(deg(\x)))*sin(deg(\y))},
				{cos(deg(\x))}
				);
				
				\draw[-latex, DarkGreen, ultra thick] (axis cs:4.0,-4,1.0) -- (axis cs:-4.0,-4,-1.0) node [circle, inner sep = 0.2pt, fill=white, above left] {\footnotesize $\mathbf{B}_{1}$};
				\draw[-latex, blue, ultra thick] (axis cs:4,-4,2.0) -- (axis cs:-4,-4,-2.0) node [circle, inner sep = 0.2pt, fill=white, above left] {\footnotesize $\mathbf{B}_{2}$};
				
				\draw[thick,dashed] (axis cs:-4,-4,0) -- (axis cs:4,-4,0);	
				\draw[thick,dashed] (axis cs:2.0,-4,-3.0) -- (axis cs:2,-4,-1);	
				
				\draw[-latex, red, ultra thick] (axis cs:2.0,-4,-3.0) -- (axis cs:1.0,-4,-1) node [circle, inner sep = 0.2pt, xshift = -0.2cm, yshift = 0.0cm] {\footnotesize $\mathbf{K}$};

				\draw[black, thick] (axis cs:-3.0,-4,0) arc [start angle=180,end angle=193,x radius=2.6cm,y radius=2.6cm] node [pos=0.5, circle, inner sep = 0.2pt, xshift = -0.3cm, yshift = 0.0cm,fill=white] {\footnotesize $\gamma_{1}$}; 
				\draw[black, thick] (axis cs:-2.0,-4,0) arc [start angle=180,end angle=205,x radius=1.8cm,y radius=1.8cm] node [pos=0.75, circle, inner sep = 0.2pt, xshift = -0.4cm, yshift = -0.1cm] {\footnotesize $\gamma_{2}$};
				\draw[black, thick] (axis cs:2.0,-4,-1.5) arc [start angle=90,end angle=125,x radius=1.0cm,y radius=1.0cm] node [pos=0.5, circle, inner sep = 0.2pt, xshift = 0.0cm, yshift = 0.2cm] {\footnotesize $\varphi_{\rm t}$};

				
			\end{axis}
	\end{tikzpicture}}
	\hspace{1em}
	\begin{tikzpicture}
        \draw (1.4,4.0)  node[above, DarkGreen]{$\varphi_{\rm t,opt,1}$};
        \draw (2.7,4.0)  node[above, blue]{$\varphi_{\rm t,opt,2}$};
		\begin{axis}[
			tick align=outside,
			tick pos=left,
			x grid style={white!69.0196078431373!black},
			xlabel={$\varphi_{\rm t} /^\circ$},
			xmin= -2.2, xmax=8.2,
			xtick style={color=black},
			y grid style={white!69.0196078431373!black},
			ylabel={$P / P_{\rm max}$},
			ymin=-0.05, ymax=1.05,
			ytick style={color=black},
			height = 5.5cm,
			width = 5.5cm
			]
			
			\addplot[DarkGreen, ultra thick, domain=-3:9, samples=50, smooth] {gaussian(2,1.5)};
			\addplot[DarkGreen, thick, dashed] table[row sep = crcr]{2 -10 \\ 2 10 \\};

			\addplot[blue, ultra thick, domain=-3:9, samples=50, smooth] {gaussian(5,1.5)};
			\addplot[blue, thick, dashed] table[row sep = crcr]{5 -10 \\ 5 10 \\};
			
		\end{axis}
	\end{tikzpicture}
	\quad
	\caption{Schematic to illustrate the toroidal response, and its geometrical intuition, for two different magnetic fields, $\mathbf{B}_{\rm 1}$ and $\mathbf{B}_{\rm 2}$. The green and blue solid lines, are the toroidal response to $\mathbf{B}_{\rm 1}$ and $\mathbf{B}_{\rm 2}$ respectively. The associated magnetic pitch angles, $\gamma_{\rm 1}$ and $\gamma_{\rm 2}$, are negative in our sign convention. The toroidal launch angle, $\varphi_{\rm t}$, which maximises the backscattered power is referred to as the optimal toroidal launch angle, $\varphi_{\rm t,opt}$. It depends on the angle between the probe beam's wavevector and magnetic field at cutoff \cite{Hillesheim:DBS_MAST:2015, Hall-Chen:beam_model_DBS:2022, Hall-Chen:mismatch:2022, Damba:mismatch:2022, Hall-Chen:mismatch:2024}. Depending on whether the pitch angle is $\gamma_1$ or $\gamma_2$ the optimal toroidal angle will differ. As such, by measuring the dependence of backscattered power on toroidal launch angle, one should be able to infer the magnetic pitch angle at the cutoff location. All other plasma properties are kept the same. Note that $\varphi_{\rm t}$ lies in the midplane and the angle shown in the figure corresponds to $\varphi_{\rm t} > 0$. Also note that the difference between $\mathbf{B}_{\rm 1}$ and $\mathbf{B}_{\rm 2}$, and therefore $\varphi_{\rm t,opt,1}$ and $\varphi_{\rm t,opt,2}$, have been exaggerated for the purpose of illustration.}
	\label{fig:ToroidalSteering}    
\end{figure*}

The rest of the paper is structured as follows. Section \ref{sec:ExperimentalSetup} details the DIII-D plasma under consideration and the data processing techniques used. We then discuss the physics basis of our proposed diagnostic in Section \ref{sec:BeamModel}. In Section \ref{sec:DeterminingMagneticPitchAngle} we use DBS to estimate the pitch angle, comparing the results with magnetics-only and MSE-constrained measurements. Lastly, in Section \ref{sec:DesignConsiderations}, we discuss how to reduce the uncertainty in the pitch angle as measured by DBS.

\section{Experimental setup}
\label{sec:ExperimentalSetup}
In this section, we describe the plasma scenario and DBS configuration used in this work.

\begin{figure*}
	\centering
	\includegraphics[width=15cm]{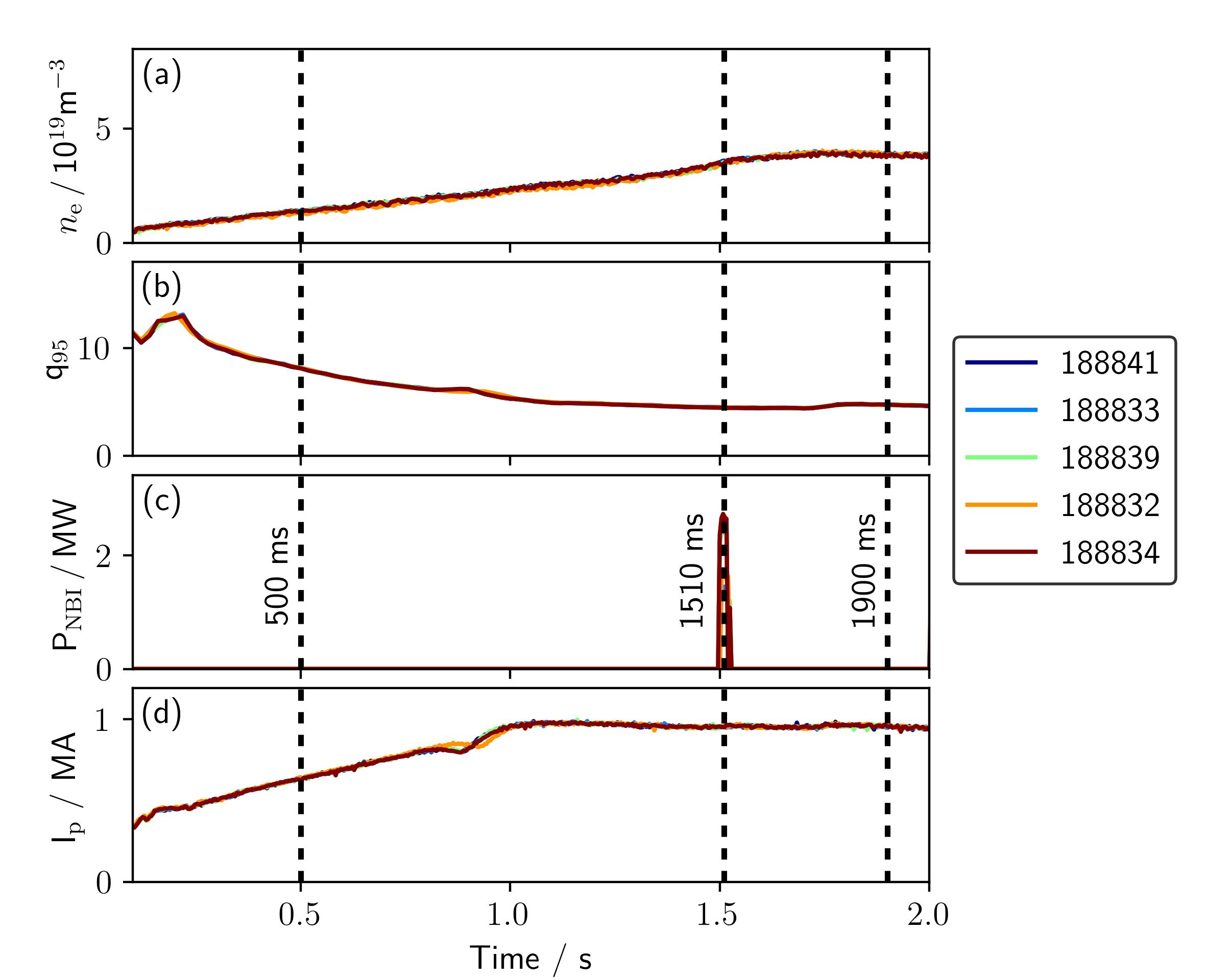}
	\caption{Plasma properties of the five repeated shots as a function of time: (a) line integrated electron density, (b) safety factor $q$ at radial coordinate $\rho = 0.95$, (c) neutral-beam power $P_{\rm NBI}$, and (d) plasma current $I_{\rm p}$. We see that these shots are indeed nearly identical. We study DBS data during plasma current ramp up, at 500 ms, and at flat top, 1510 ms; these two times are marked with dotted lines. While eleven repeated shots were analysed in this paper, here we only show the five with significant DBS signal; while the other six shots had nearly identical plasma properties, the DBS signals across all frequency channels were negligible due to prohibitively high mismatch attenuation \cite{Hall-Chen:mismatch:2022, Damba:mismatch:2022}.}
	\label{fig:ScopeFig}
\end{figure*}

As we will see in Section \ref{sec:DeterminingMagneticPitchAngle}, inferring the magnetic pitch angle with DBS relies on simultaneous measurements at the same radial location but with different toroidal launch angles, enabling the backscattered power as a function of toroidal launch angle to be calculated. However, such simultaneous measurements require specialised hardware or steering the DBS launch optics during a plasma discharge, which is currently not possible on DIII-D. Instead, we repeat the same shot, varying the DBS toroidal launch angle from shot to shot. These shots are designed to be as similar to each other as possible, with the same density, shaping, q-profile, and plasma current. We used a series of eleven repeated shots conducted on the DIII-D tokamak \cite{Buttery:DIIID:2023}; these same shots, at 1900ms, were previously used for validating our model of mismatch attenuation \cite{Damba:mismatch:2021, Damba:mismatch:2022, Hall-Chen:mismatch:2022}. These shots do indeed have nearly identical plasma properties, see Figure \ref{fig:ScopeFig}. The electron density was measured with Thomson scattering and processed with the OMFITprofiles module \cite{logan2018omfitprofiles}. Magnetic equilibrium profiles were obtained from Equilibrium Fitting (EFIT) \cite{Lao:EFIT:1985, Appel:EFIT:2006}. MSE measurements of the magnetic pitch angle were only available at 1510 ms, during a neutral beam injection blip, see Figure \ref{fig:ScopeFig}.

We use the DBS system located at DIII-D's 240$^\circ$ port. DBS240 has eight frequency channels: 55.0 GHz, 57.5 GHz, 60.0 GHz, 62.5 GHz, 67.5 GHz, 70.0 GHz, 72.5
GHz, and 75.0 GHz. The probe beam can be steered toroidally and poloidally with a mirror. This steering was calibrated by setting up a target board in the tokamak and shining a laser beam through the quasioptical system. The laser beam formed a spot with radius of approximately 0.5 cm on the target board, and the board was 1.2 m from the mirror. As such, we estimate the toroidal steering precision, $\Delta \varphi_{t, err} \approx 0.3 ^\circ$. Further details of this DBS system, such as its quasioptics and gain, are available in previous work \cite{Rhodes:DBS:2018}.

\begin{figure*} 
	\centering
	\includegraphics[width=15cm]{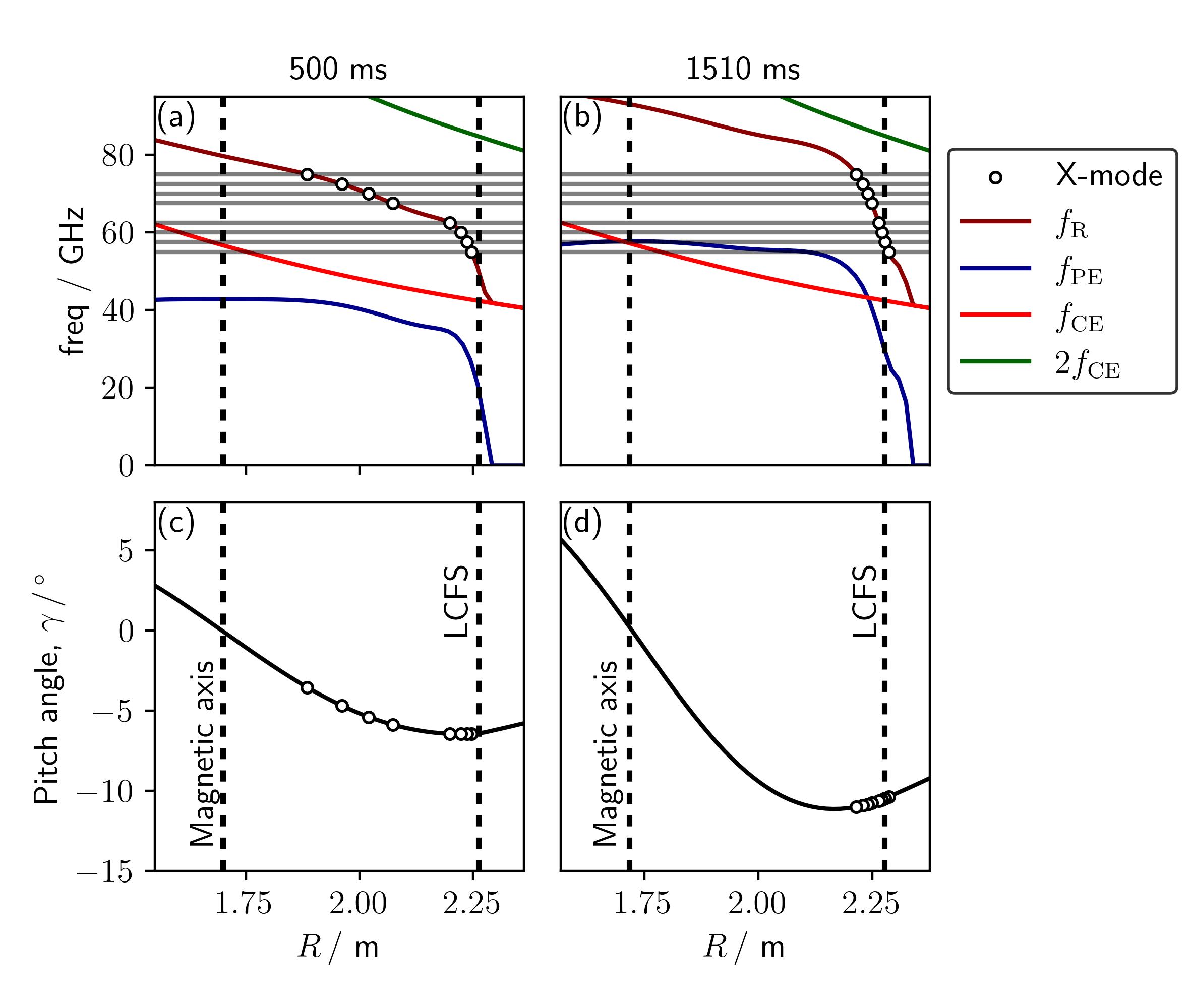}
	\caption{Properties of the plasma at midplane for the time slices 500 ms (left) and 1510 ms (right). The experiments were carried out in X-mode polarisation for the set of frequencies: 55.0 GHz, 57.5 GHz, 60.0 GHz, 62.5 GHz, 67.5 GHz, 70.0 GHz, 72.5 GHz, and 75.0 GHz, as indicated by gray horizontal lines. Points with white centers are the cutoff positions calculated for launch at normal incidence on the midplane. Characteristic frequencies at midplane (top): $f_{\rm R}$ is the right-hand cutoff frequency, $f_{\rm PE}$ is the plasma frequency and the cutoff frequency for O-mode, $f_{\rm CE}$ is the cyclotron frequency, $2 f_{\rm CE}$ is the second harmonic of the cyclotron frequency. Pitch angle $\gamma$ at the midplane (bottom). We chose 1510 ms as the main time slice of study because it is the only time when MSE measurements are available. At this time, X-mode measurements were concentrated at the edge. We also analyse data from 500 ms, when the cutoff positions of the frequencies were deeper in core, probing a wider range of pitch angles. }
	\label{fig:equilibrium500ms1510ms}
\end{figure*}

Across the eleven repeated shots, the DBS poloidal launch angle was fixed at $-11.4^\circ$ while its toroidal launch angle was varied from $-6.5^\circ$ to $10.1^\circ$. The sign convention for these angles is defined in equation (\ref{eq:launchK}). The DBS polariser was set to couple with X-mode during flat top, that is, when the plasma current is constant. In this paper, we focus on DBS measurements at 1510 ms, during flat top, the only time when MSE measurements were available. Additionally, to investigate how DBS might make pitch angle measurements in the core, we also study DBS data during ramp up (500 ms). At this time, the density was low enough for X-mode to access the core, unlike during flat top: all eight DBS channels' cutoff locations were near the edge, see Figure \ref{fig:equilibrium500ms1510ms}. Since we expect magnetics-only reconstruction to be more accurate at the edge and less accurate in the core, having DBS measurements that span the edge to core is critical. At 500 ms, during ramp up, the cutoff locations do indeed span this range, Figure \ref{fig:equilibrium500ms1510ms}. Unfortunately, some shots had slightly higher or lower densities at a given time during ramp up.  We wanted to find times in the ramp up where the shots had the same density; we thus selected times between 460 ms and 580 ms, with seven of the eleven shots selected at 500 ms. The edge pitch angle at 500 ms was $5^\circ$ lower than that at flat top, as determined by magnetics. Since the DBS polariser was adjusted to couple to X-mode in the edge during flat top, we expect some spuriously coupled O-mode during ramp up. Nonetheless, we expect the spurious coupling to be low. Indeed, the backscattered signal from the O-mode is not visible in the backscattered power spectra. 
Like traditional DBS, our approach requires that the DBS beam frequency avoids the electron cyclotron frequency and its harmonics to prevent resonant absorption (see Section 3.3.1 of \cite{hillesheim2012thesis}). The DBS hardware can also be susceptible to damage from stray electron-cyclotron radiation (see Section 3.3.4 of \cite{hillesheim2012thesis}). Finally, large density profiles at the edge, such as those in H-mode MAST-U plasmas \cite{Shi:DBS:2023}, might limit core access. While the plasma scenario used in this paper avoids these issues, it is important to consider them when applying this technique to other devices and scenarios.

Our analysis of DBS data follows routine techniques employed by previous work \cite{Hillesheim:DBS_MAST:2015, Damba:mismatch:2022}. We perform a short-time Fourier transform, with the following parameters: the data was sampled at 5 MHz, 4096 points were used for each Hann window (0.82 ms per window), the overlap fraction between the windows is 0.25, and lastly, we smooth the time-averaged spectrogram with an unweighted moving average that is 20 kHz wide. We fit the resulting spectrogram with a two-peak Student's T model, a Lorentzian-Gaussian mixture, to remove spurious contributions near 0 MHz and to calculate the backscattered power from the Doppler-shifted peak. This two-peak model approach was used and further explained in previous work \cite{pratt2022comparisonExB,Pratt:OMFIT:2024, pratt2024phdthesis}. We used the backscattered power from every fit in a 10 ms interval to calculate the mean backscattered power and the associated error bar. This analysis was performed in the DBS module \cite{Pratt:OMFIT:2024} in the OMFIT framework \cite{Meneghini:OMFIT:2013, Meneghini:OMFIT:2015}. 

We aim to infer the magnetic pitch angle by comparing the difference in the measured and predicted mismatch attenuation as a function of toroidal launch angle. Hence, we need to calculate the mismatch attenuation a priori, rather than empirically, which we shall describe in the next section.

\section{Calculating mismatch attenuation with the beam model of DBS}
\label{sec:BeamModel}
In this section, we describe and motivate the quantitative model used for calculating the mismatch attenuation; this model is predictive and no fitting of experimental measurements is required. We begin by explaining our choice of beam tracing and the reciprocity theorem to interpret DBS, which we call the beam model. We then briefly summarise the assumptions employed in this beam model and the simplifications used in this paper. Finally, we explain the physical intuition behind the quantitative description of mismatch attenuation.

While DBS seeks to measure turbulent density fluctuations and flows, there are many other complicated contributions to the backscattered signal, which we call instrumentation effects \cite{Hall-Chen:beam_model_DBS:2022, Gusakov:scattering_slab:2004, Frank:spectrum_resolution:2023, RuizRuiz:slab:2024}. Correcting for these instrumentation effects is key for quantitatively interpreting DBS measurements. As such, it is crucial to calculate the electric field of the DBS probe beam. For example, regions of the plasma where the probe beam's electric field is larger will have more scattering, even if the turbulent density fluctuations have the same amplitude. Ray tracing is typically used to calculate the probe beam's trajectory through the plasma and the associated wavevector. However, the beam bends strongly at the cutoff, which causes rays to cross. At these crossings, known as caustics, the calculated electric field diverges. Since the region near the cutoff is where we expect the strongest contribution to backscattering \cite{Hirsch:DBS:2001}, the usefulness of ray tracing is thus limited. While caustics can, in principle, be resolved \cite{Tracy:ray_tracing:2014, Lopez:MGO:2022, Lopez:MGO:2023}, this is complicated in practice and beyond the scope of this paper. Another approach is to solve Maxwell's equations directly to find the backscattered power; such full-wave simulations \cite{Hillesheim:2D_fullwave:2012, Happel:DBS_synthetic:2017, Frank:spectrum_resolution:2023} have contributed significantly to our understanding of DBS. Unfortunately, there are several drawbacks, such as requiring one to assume a particular instantiation of turbulence. The drawback most relevant to this paper is their computational intensiveness, which typically limit simulations to 2D geometries in practice. Since mismatch attenuation is a 3D effect, using full-wave simulations is likely not the best approach. Instead, we use beam tracing to determine the probe beam's electric field. By going to an order higher than standard ray-tracing, we evolve a Gaussian envelope around a single guiding central ray instead of tracing a bundle of rays. Since there is only one ray, there are no crossings and thus no caustics. Beam tracing is a well-established technique \cite{Pereverzev:Beam_Tracing:1996, Pereverzev:Beam_tracing:1998, Poli:Torbeam:2001} and has been shown to be applicable near cutoffs \cite{Maj:Beam_Tracing:2009}. We use beam tracing to determine the probe beam's electric field and the reciprocity theorem \cite{Piliya:reciprocity:2002, Gusakov:scattering_slab:2004} to determine the linear scattered electric field; we call this approach the beam model of DBS \cite{Hall-Chen:beam_model_DBS:2022}, which has been experimentally validated on both conventional \cite{Hall-Chen:mismatch:2022, Damba:mismatch:2022} and spherical \cite{Hall-Chen:mismatch:2024} tokamaks. This model is implemented in an open-source beam-tracing code, Scotty \cite{Hall-Chen:Scotty:2022}. The beam model can use any beam tracer, such as TORBEAM \cite{Poli:Torbeam:2001}, should the appropriate post-processing techniques be applied. In the rest of this section, we introduce the beam model of Doppler backscattering and summarise our model's assumptions and key relevant results. Throughout the rest of this paper, we then use this model to predict the effect of magnetic pitch angle on backscattered power. 

In standard ray tracing, we assume that the beam's wavelength, $\lambda$, is much smaller than the inhomegeneity scale, $L$. In beam tracing, we introduce an intermediate length scale, such that
\begin{equation}
\label{eq:BeamTracingOrdering}
	\frac{\lambda}{W} \sim \frac{W}{L} \ll 1 ,
\end{equation}
where $W$ is the characteristic width of the electric field's Gaussian envelope. This ordering arises from taking the Rayleigh length of the Gaussian beam to be on the order of $L$. One can then derive, from Maxwell's equations, the evolution equations of a Gaussian beam \cite{Pereverzev:Beam_tracing:1998, Hall-Chen:beam_model_DBS:2022}. With these equations and the initial beam wavevector, position, Gaussian width, and wavefront curvature, it is then an initial-value problem to solve a system of ordinary differential equations and calculate the probe beam's electric field inside the plasma. We then Fourier analyse the electron density fluctuations from which the beam scatters, $\delta n_{\rme} \left( \mathbf{r}, t \right)$, in a tube-like volume containing this Gaussian beam and where the electric field is essentially zero on the surface bounding this volume. We assume that these fluctuations have small gradients parallel to the magnetic field and large gradients perpendicular to it \cite{Catto:GK:1978, Frieman:GK:1982}. As such, we neglect the component of the turbulent wavevector parallel to the magnetic field and only Fourier analyse in the two directions perpendicular to it. Consequently, the density fluctuations can be expressed as $\delta \tilde{n}_\rme \left( \mathbf{k}_{\rm \perp}, u_{\rm \parallel}, \omega \right)$, where $u_{\rm \parallel}$ is the location along the magnetic field lines (measured in units of lengths), $\mathbf{k}_{\rm \perp}$ is the fluctuation wave-vector, and $\omega$ has contributions from both the turbulent frequency in the plasma's frame and the Doppler shift due to plasma flow. We assume that the amplitude of these fluctuations is much smaller than both the equilibrium electron density and the threshold for non-linear multi-scattering events \cite{Gusakov:non-linear:2005, Krutkin:non-linear:2019}. The beam model of DBS, as implemented in Scotty, is thus a quantitative synthetic diagnostic that is capable of calculating the linear backscattered signal in general geometry, including backscattering away from the cutoff.

For the purposes of this paper, we make three simplifications to the beam model. While these simplifications are not strictly necessary, they help make the model easier to understand without having a significant impact on results from DIII-D. The first simplification is assuming that statistical properties of turbulent fluctuations are unvarying in the region where the beam propagates. As such, we avoid having to use the correlation function, and the backscattered spectral density \cite{Hall-Chen:beam_model_DBS:2022}, $p_{\rm r} (\omega)$, is given by 
\begin{equation} \label{eq:PowerReceived}
	\eqalign{
		p_{\rm r} (\omega)
		=
		C
		\int 
		F_{\rm i} F_{\rm m}
		\left|
		\delta \tilde{n}_{\rme} (\mathbf{k}_{\rm \perp} (l), u_{\rm \parallel} (l), \omega)
		\right| ^2
		\ \rmd l .
	}
\end{equation}
Here $C$ is a known constant, $l$ is the arc length along the central ray, $\delta \tilde{n}_{\rme} $ is the wavenumber spectrum of the turbulent density fluctuations, $F_{\rm m}$ accounts for mismatch attenuation, and $F_{\rm i}$ is the product of the other instrumentation functions. Note that $F_{\rm i}$ and $F_{\rm m}$ depend on properties of the probe beam, such as wavenumber, beam width, and wavefront curvature. The wavevector of the turbulent density fluctuations responsible for scattering is given by the Bragg condition $\mathbf{k}_{\rm \perp} = - 2 \mathbf{K}$ at every point along the ray. The component of $\mathbf{k}_{\rm \perp}$ perpendicular to $\mathbf{K}$ contributes to wavenumber resolution, which has already been integrated over to obtain equation (\ref{eq:PowerReceived}) \cite{Hall-Chen:beam_model_DBS:2022}. However, we remark that a poor wavenumber resolution would widen the Doppler-shifted peak and therefore impair our ability to extract the Doppler-shifted component.

The mismatch attenuation is a Gaussian function in mismatch angle, given by
\begin{equation} \label{eq:FilterMismatch}
	F_{\rm m}
	=
	\exp\left[
	- 2
	\frac{\theta_{\rm m}^2}{\left( \Delta \theta_{\rm m} \right)^2}
	\right] .
\end{equation}
Here $\Delta \theta_{\rm m}$ is the $1 / \rme^2$ width of the mismatch attenuation, which we call the mismatch tolerance. This mismatch tolerance is a function of beam wavenumber, width, wavefront curvature, and properties of the plasma equilibrium, specifically, the curvature and shear of the magnetic field. Mismatch tolerance's explicit dependence on these variables is given in previous work \cite{Hall-Chen:beam_model_DBS:2022}. In this paper, we calculate mismatch tolerance with the Scotty code. The mismatch tolerance is inversely proportional to beam wavenumber and has a complicated dependence on beam width and wavefront curvature. The physical intuition is that the narrower or more curved the beam in real space, the larger the spread of wavevectors in k-space, which makes it easier for one of these wavevectors to meet the Bragg condition. This wavefront curvature is modified by the magnetic field's curvature and shear. The second simplification is neglecting the contribution of the other instrumentation functions, $F_{\rm i}$, which account for effects such as larger electric fields due to smaller beam widths \cite{Hall-Chen:beam_model_DBS:2022, RuizRuiz:slab:2024}. While $F_{\rm i}$ typically plays an important role in interpreting DBS, it does not vary significantly when the DBS toroidal launch angle is changed for the shots studied. By way of due diligence, we tried including $F_{\rm i}$ in our analysis but found that it made no material difference for the shots studied. This is a result of the cutoff location and beam wavenumber at cutoff remaining roughly constant during the toroidal sweep, as we show later in Section \ref{subsec:ToroidalAngleSweep}. The third and final simplification is assuming that the backscattered signal comes entirely from the nominal cutoff location, that is, the point along the central ray where $K$ is minimised. We found that the line-integrated signal yields effectively the same result as assuming the signal comes entirely from the cutoff, as such, we only show the latter in this paper.

Having explained the physical model behind mismatch attenuation in this section, we move on to show how we can infer the magnetic pitch angle from matching.

\section{Inferring the magnetic pitch angle with DBS}
\label{sec:DeterminingMagneticPitchAngle}
Our proposed method of using DBS to give a relative but localised measurement of the magnetic pitch angle is illustrated in Figure \ref{fig:flowchart} and described in the rest of this paragraph. We seek to have several simultaneous DBS measurements at the same radial location, with each measurement having different mismatch angles only. This is achieved by varying the toroidal launch angle, which in turn varies the mismatch angle at cutoff. The backscattered signal is maximised when the mismatch at cutoff is zero, that is, when the probe beam's wavevector is locally perpendicular to the magnetic field \cite{Hillesheim:DBS_MAST:2015, Damba:mismatch:2022}. We call the corresponding toroidal launch angle the optimal launch angle. Scotty's ability to predict dependence of backscattered power on toroidal launch angle was established in previous work \cite{Hall-Chen:beam_model_DBS:2022, Hall-Chen:mismatch:2022, Hall-Chen:mismatch:2024}. Scotty requires the electron density and an estimate of the equilibrium magnetic field as input, which are measured by Thomson scattering and magnetics, respectively. We now work in reverse; we attempt to determine the pitch angle by comparing the empirical dependence of backscattered power on toroidal angle with that calculated from Scotty. 
\begin{figure*} 
	\centering
	\includegraphics[width=15cm]{flowchart.jpg}
	\caption{Workflow of inferring to magnetic pitch angle from DBS measurements, as used in this paper. We show experimental measurements (blue boxes), codes (green boxes), Scotty's input parameters (orange boxes), intermediate parameters (red boxes), and the output parameter (purple box).}
	\label{fig:flowchart}
\end{figure*}
This requires the cutoff location and wavenumber at cutoff to remain constant when the toroidal launch angle is varied, which we show in subsection \ref{subsec:ToroidalAngleSweep}. To elucidate how pitch angle affects the backscattered power as a function of toroidal angle, we artificially scale the magnetic pitch angle in our beam-tracing simulations, subsection \ref{subsec:DbsMeasurementsPitchAngleScaling}. Finally, using DBS measurements of repeated shots from DIII-D, we show how the pitch angle at a given spatial location may be inferred, comparing this DBS-measured pitch angle with that from magnetics and MSE, subsection \ref{subsec:InferPitchAngle}. This final subsection is the crux of this paper.

\subsection{Toroidal launch angle sweep}    
\label{subsec:ToroidalAngleSweep}
\begin{figure*}
	\centering
	\includegraphics[width=15cm]{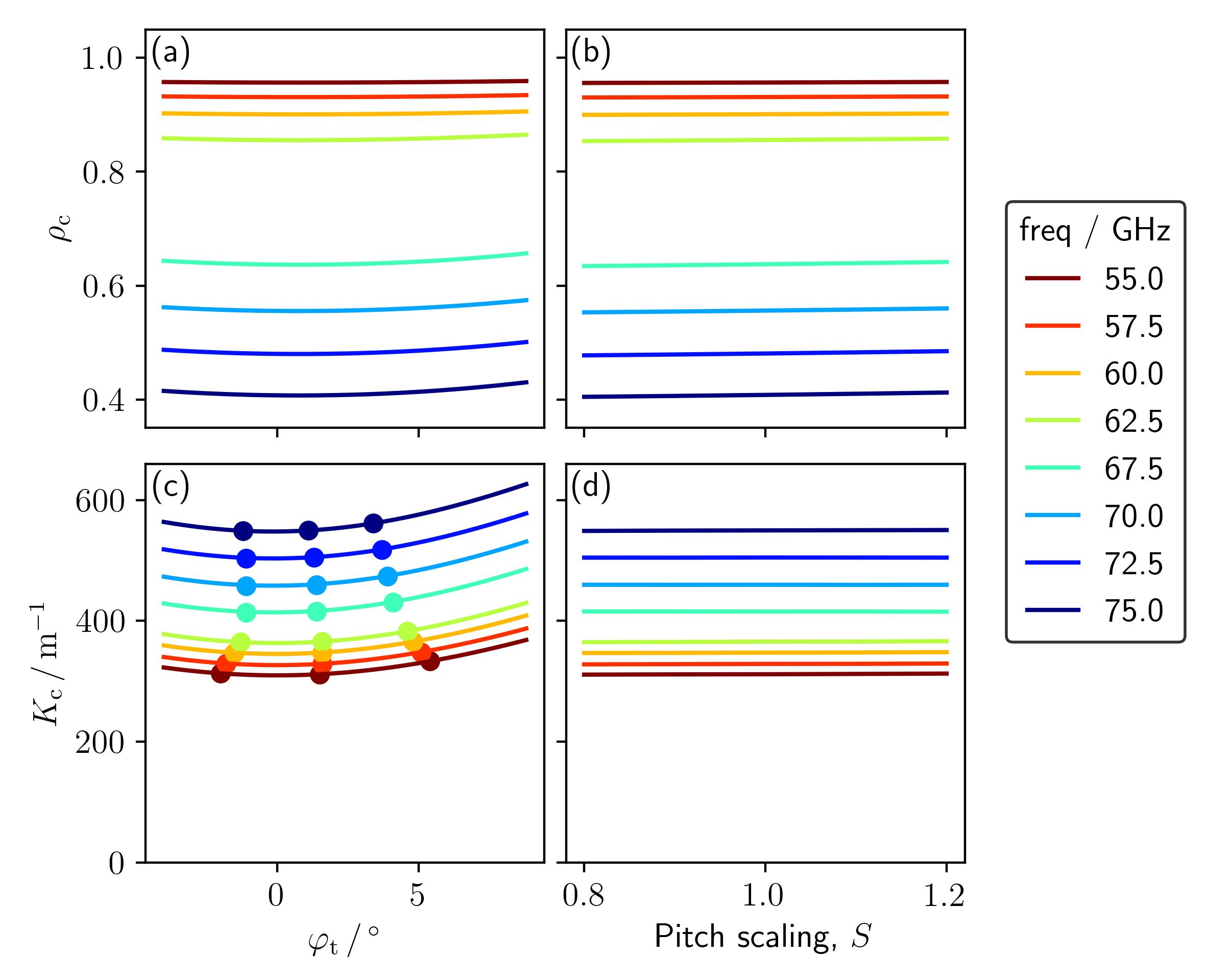}
	\caption{Radial location of cutoff locations, $\rho_{\rm c}$, (top) and wavenumbers at cutoff, $K_{\rm c}$, (bottom) at 500 ms as a function of toroidal launch angle, $\varphi_{\rm t}$, (left) and as a function of pitch angle scaling, $S$, (right). The latter is discussed in detail in subsection \ref{subsec:DbsMeasurementsPitchAngleScaling}. The equilibrium from shot 188839 was used for these beam-tracing simulations. The DBS was set to X-mode and poloidal launch angle $\varphi_{\rm p} = -11.4^\circ$. The sign convention for toroidal and poloidal angles is given in equation (\ref{eq:launchK}). Here $\rho$ is the square root of normalised toroidal flux. The three dots on each line (c) demarcate the following: the center dot is the optimal toroidal angle at which we get maximum backscattered power, the leftmost and rightmost dots are at toroidal angles such that the backscattered power is $1/\rme^2$ of the maximum backscattered power. $K_{\rm c}$ is approximately constant in this range of toroidal launch angles (c) and does not depend on pitch angle scaling (d). The radial coordinate $\rho_{\rm c}$ of the cutoff location has little dependence on toroidal launch angle (a) and no dependence on pitch angle scaling (b). At 1510ms, the dependence of $K_{\rm c}$ and $\rho_{\rm c}$ on $\varphi_{\rm t}$ and $S$ weakens further (not shown).}
	\label{fig:CutoffWavenumberVSToroidalScaling}
\end{figure*}

To greatly simplify interpretation of the dependence of backscattered power on toroidal angle, we seek to have the same cutoff location, $\rho_c$, and probe beam wavenumber at cutoff, $K_c$. The cutoff is where most of the power is backscattered, hence it determines the properties of the plasma that we are actually measuring. Moreover, the turbulence spectrum depends heavily on wavenumber \cite{Schekochihin:spectrum:2008, Schekochihin:spectrum:2009, Barnes:spectrum:2011, Pratt:spectrum:2023} as does a part of the DBS instrumentation function \cite{Hall-Chen:beam_model_DBS:2022}. We find that the radial location of the cutoff position and wavenumber at cutoff are effectively constant as toroidal launch angle is varied, see Figure \ref{fig:CutoffWavenumberVSToroidalScaling}, see Figure \ref{fig:wavenumberInfluence} for a discussion on the negligible effect of $K_{\rm c}$ on toroidal response for the shots studied. Since we are measuring the same pitch angle, and locations with similar turbulence characteristics, we attribute the toroidal response to be predominantly from mismatch attenuation. 


\subsection{Scaling the magnetic pitch angle to understand its effect on DBS}
\label{subsec:DbsMeasurementsPitchAngleScaling}

\begin{figure}
	\centering
	\includegraphics[width=8cm]{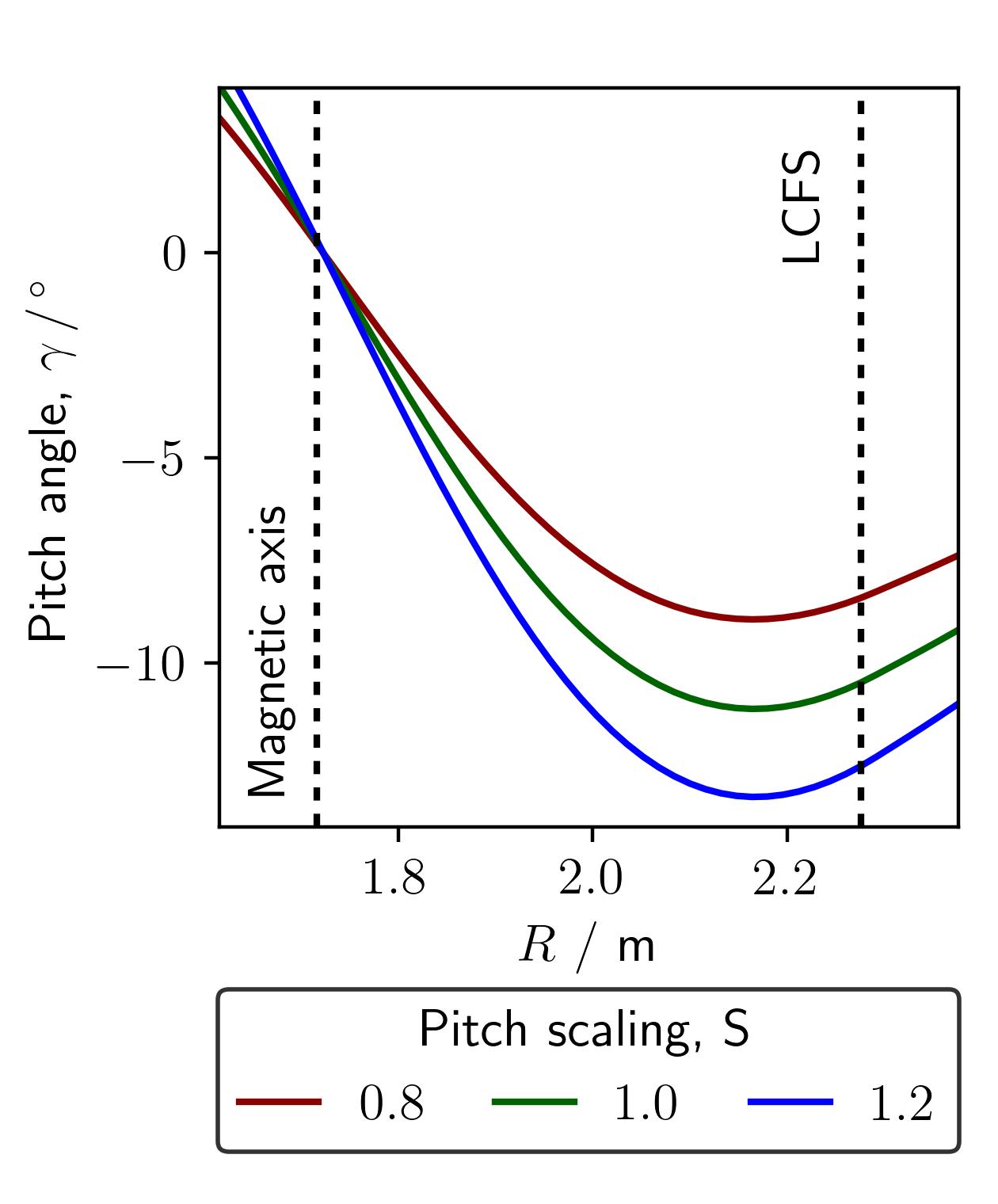}
	\caption{Magnetic pitch angles, given by $\gamma = \tan^{-1} \left(B_{\rm p} / B_\zeta \right)$, at the midplane for shot 188839 at 1510 ms. The green line shows $\gamma$ as calculated from magnetics-only EFIT. We use the Bateman transformation \cite{bateman1977magnetohydrodynamic} to globally and uniformly scale the plasma current, as obtained from EFIT, up and down by 20\% without changing the shape of the flux surfaces. The corresponding pitch angles are scaled accordingly (red and blue lines). }
	\label{fig:BatemanScaling}
\end{figure}

We seek to isolate the effect of the magnetic pitch angle on the toroidal response. To that end, we scale the magnetic pitch angle, at all points in the plasma, near-uniformly by a scaling factor, $S$. We require a method of scaling that maintains the same flux surface shapes and ensures that the new scaled equilibrium is still a solution of the Grad-Shfranov equation. Hence, we use the Bateman transformation \cite{bateman1977magnetohydrodynamic}; for small scalings, this is effectively a global near-uniform scaling of current density and therefore pitch angle, as shown in Figure \ref{fig:BatemanScaling}. 
Using the beam-tracer Scotty, we confirm that scaling the pitch angle does not affect the cutoff location or wavenumber at cutoff, see Figure \ref{fig:CutoffWavenumberVSToroidalScaling}. This transformation scales the mismatch angle at all points along the beam, but does not significantly change any other aspect of the beam.

\begin{figure*}
	\centering
	\includegraphics[width=15cm]{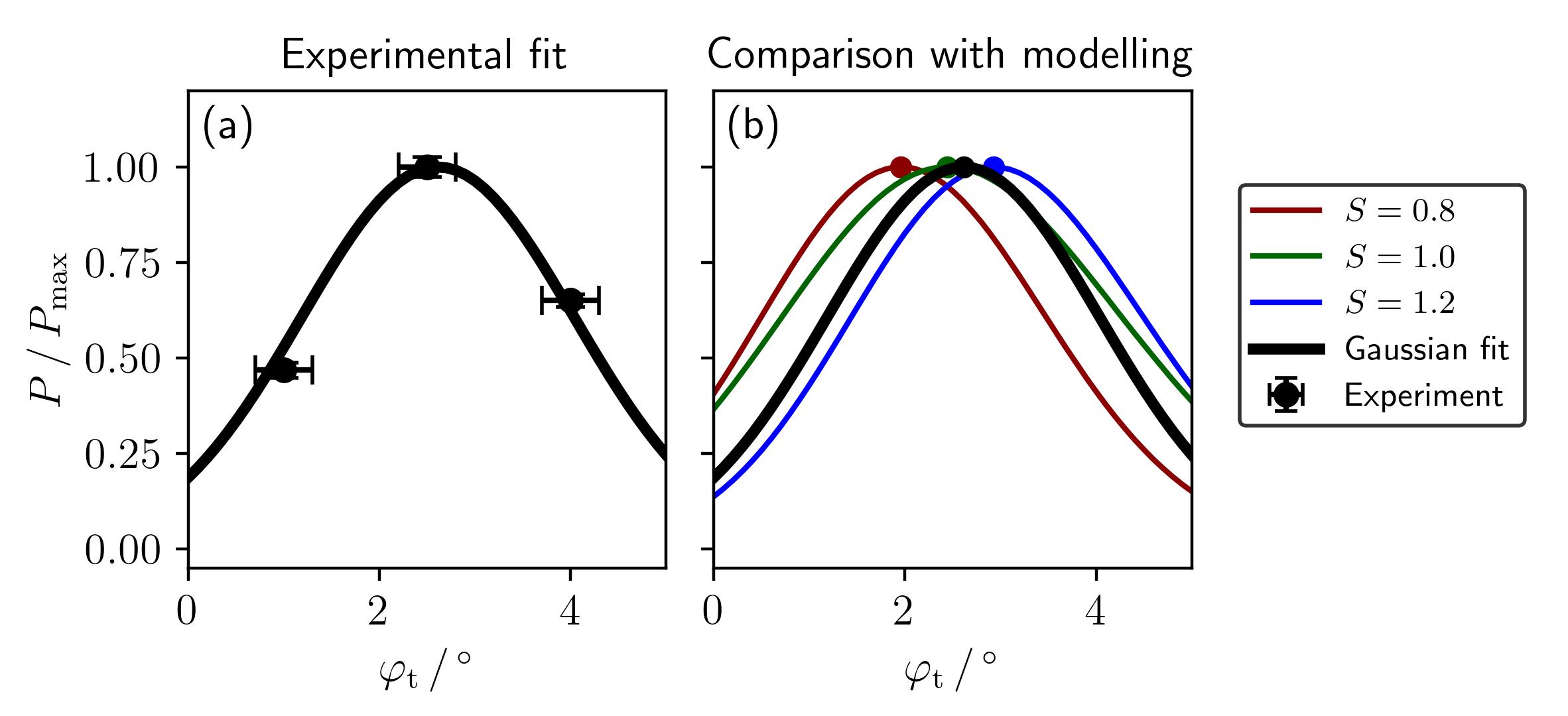}
	\caption{Experimentally measured backscattered power (a) and simulated effect of pitch-angle scaling, $S$, on backscattered power (b), both as functions of toroidal launch angle, $\varphi_{\rm t}$. Data and simulations are from the 67.5 GHz channel at 1510 ms. Vertical experimental error bars are calculated using the same method as previous work \cite{Hillesheim:DBS_rotation:2015} while horizontal error bars indicates the toroidal steering precision, see Section \ref{sec:ExperimentalSetup}. The black line is a Gaussian fit to the experimental data obtained from DBS. The covariance of the Gaussian peak's location later gives the error in the measured magnetic pitch angle. }
    
	\label{fig:gaussians}
\end{figure*}

\begin{figure*}
	\centering
	\includegraphics[width=15cm]{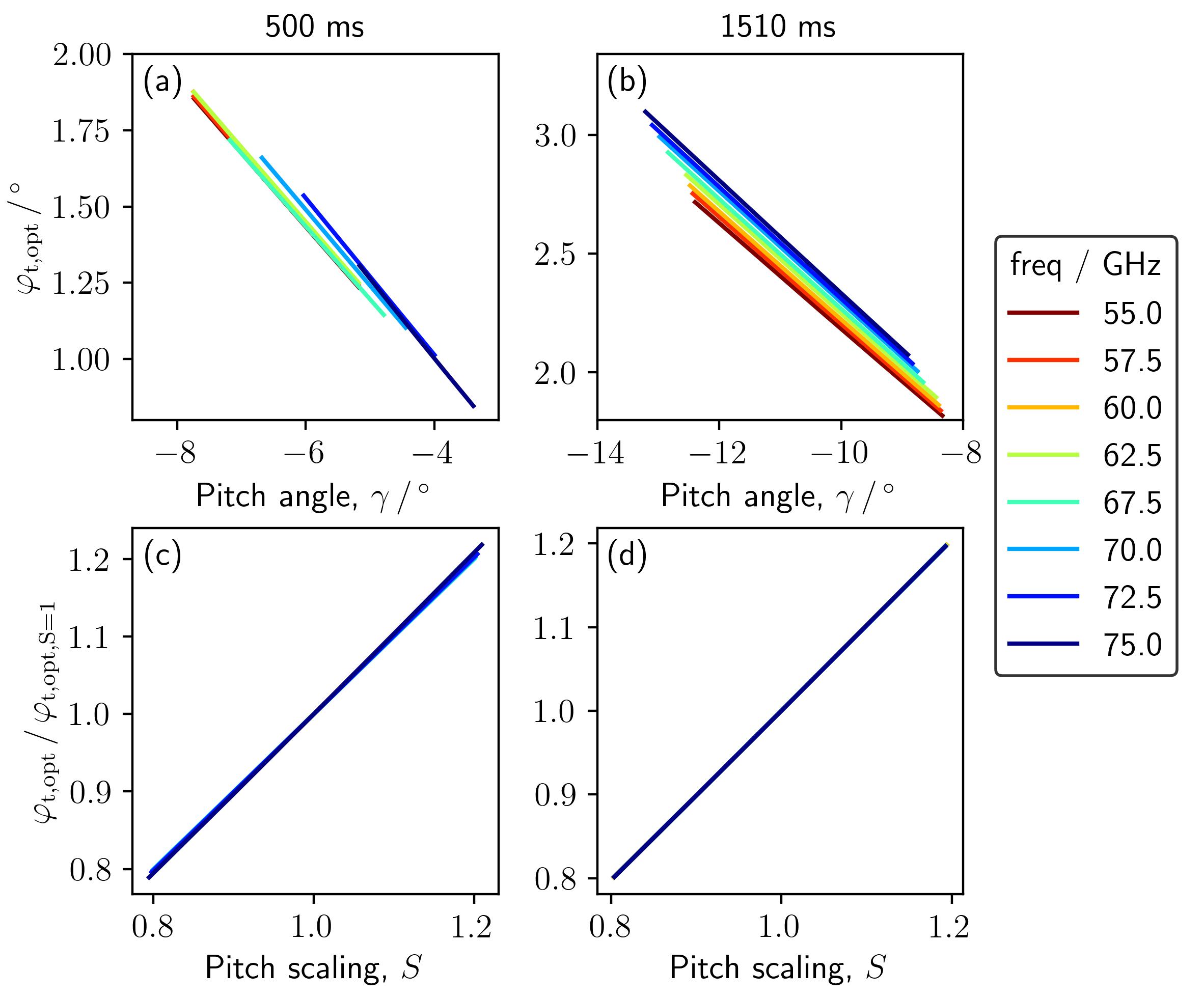}
	\caption{How scaling the pitch angle, $S$, affects the optimal toroidal angle, $\varphi_{\rm t, opt}$, for 500 ms (left, a and c) and 1510 ms (right, b and d). Variation of optimal toroidal angle with pitch angle, $\gamma$, at cutoff (top, a and b). There is a clear linear relationship between the predicted optimal toroidal angle and the pitch angle at cutoff. Normalised optimal toroidal angle $\varphi_{\rm t, opt}/ \varphi_{\rm t, opt, S=1}$ against pitch angle scaling $S$ (bottom, c and d), note that the gradients are the same for all frequencies and both cases. A given percentage change in the pitch angle leads to the same percentage change in optimal toroidal launch angle.}
	\label{fig:TorOptimNormComb}
\end{figure*}

Having established the suitability of the Bateman transformation for our purposes, we proceeded to use it to scale pitch angle to be 20\% larger and smaller than magnetics-only reconstruction, in steps of 10\%. We used Scotty to simulate the effect of this scaling on the toroidal response. As one might expect, the optimal toroidal angle increases with the pitch angle, see Figure \ref{fig:gaussians}.
For each frequency and pitch angle scaling, we used Scotty simulations to determine the optimal toroidal angle --- the angle at which backscattered power is maximised --- which happens when the mismatch angle at cutoff is zero. For both core and edge measurements in the shots studied, we found that the optimal toroidal launch angle varies linearly with pitch angle scaling, see Figure \ref{fig:TorOptimNormComb}. That is, a given percentage increase in the pitch angle leads to the same percentage increase in the optimal toroidal launch angle. Interestingly, we find that the constant of proportionality is the same for all channels at both times. Whether this is also the case for other devices, and why, would be an interesting topic for future study.
Crucially, this linear relationship implies that measuring the optimal toroidal launch angle yields information about the magnetic pitch angle, which can then be exploited to use DBS as a pitch angle diagnostic. In the next subsection, we show one possible approach of determining the pitch angle with DBS.

\subsection{Method of inferring the magnetic pitch angle with DBS}
\label{subsec:InferPitchAngle}        
In the previous subsection, we established that there is a linear relationship between pitch angle and the optimal toroidal launch angle for the shots studied. In this subsection, we reverse the process, using the measured optimal toroidal angle to infer the pitch angle at cutoff.

\begin{figure}
	\centering
	\includegraphics[width=8cm]{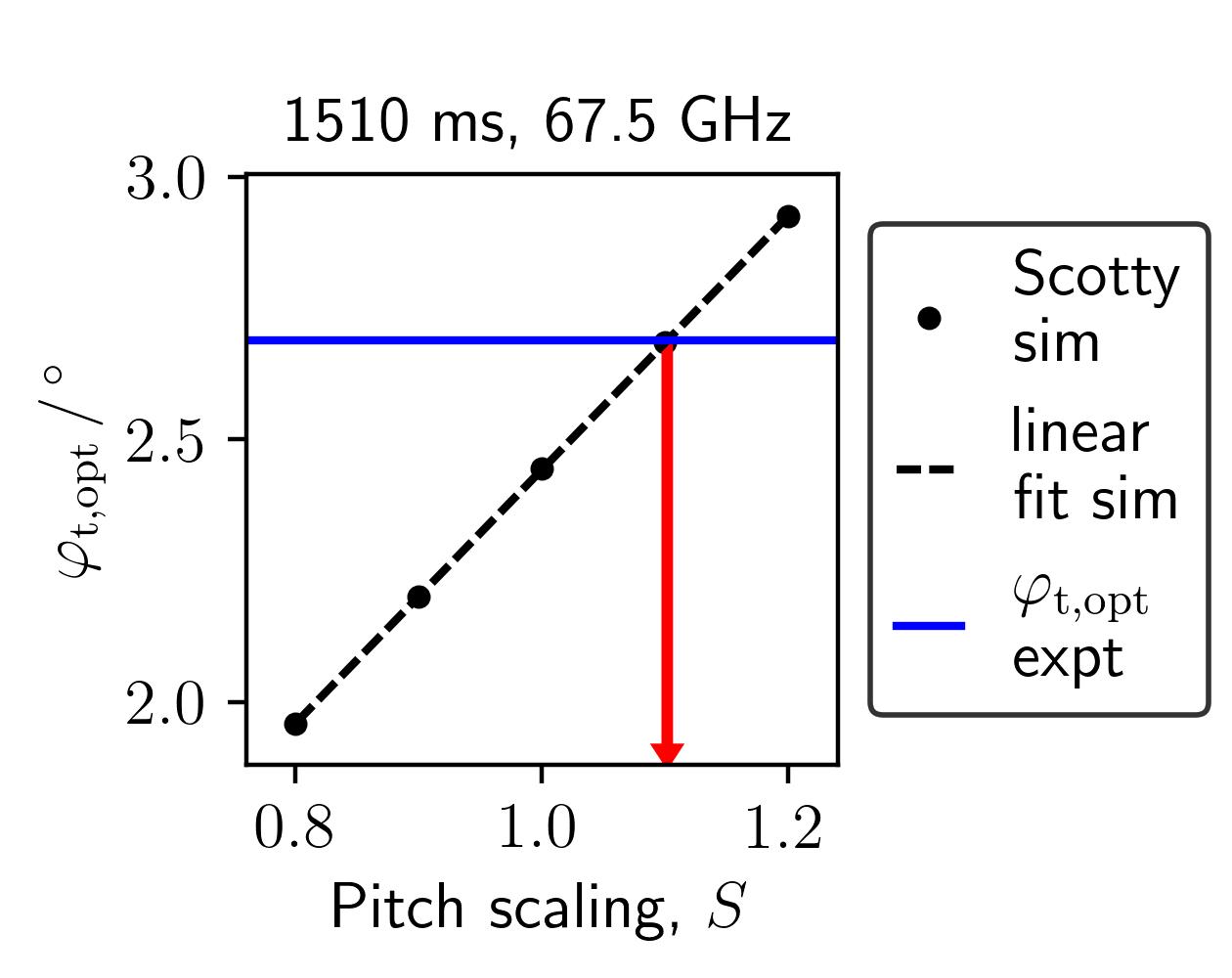} 
	\caption{We use Scotty to calculate the dependence of optimal toroidal launch angle on pitch angle scaling, $S$, and linear fit of these results (black dashed line). We then find the experimental optimal toroidal launch angle (blue line), $\varphi_{\rm t, opt}$, by fitting a Gaussian to the dependence of DBS power on toroidal angle, see black line in Figure \ref{fig:gaussians}. Finally, we determine the pitch angle scaling $S$ (red downward-pointing arrow) by finding the intersection between the experimentally measured optimal launch angle and the linear fit of simulated optimal toroidal angles. The pitch angle at the $67.5$ GHz cutoff location is then estimated by scaling the initial equilibrium with the scaling factor obtained.}
	\label{fig:CurrentScaling67.5}
\end{figure}

First, for every DBS frequency, we systematically vary the pitch angle scaling and run Scotty simulations to determine the optimal toroidal launch angle as a function of scaling. Secondly, we determine the empirically measured optimal toroidal launch angle by fitting a Gaussian to experimental measurements of backscattered power as a function of toroidal launch angle, see black line in Figure \ref{fig:gaussians}. The peak of the Gaussian fit gives the optimal launch angle. Thirdly, we compare the optimal angles calculated from the preceding two steps, determining the pitch angle scaling which would give the measured optimal toroidal launch angle. This particular scaling tells us how much our equilibrium needs to be scaled for simulations and experiments to agree. Finally, we repeat the aforementioned three steps for each frequency channel, getting the pitch angles at the cutoff positions. This process is shown in Figure \ref{fig:CurrentScaling67.5}. It is worth noting that the precision of toroidal steeing is $\sim 0.3^\circ$ (Section \ref{sec:ExperimentalSetup}) while the change in optimal toroidal angle is $\sim 1.0^\circ$ when the pitch angle scaling is changed from $S=0.8$ to $1.2$. We later show, with a simple Monte Carlo analysis, that the optimal toroidal angle can be determined with a precision of around $0.15^\circ$, see subsection \ref{subsec:accuracy}. Moreover, we show in Section \ref{sec:DesignConsiderations} that changing the launch beam properties (subsection \ref{subsec:InitialBeamConditions}) and launch poloidal angle (subsection \ref{subsec:OptimPolAngle}) can further improve the precision in determining the optimal toroidal angle.

\begin{figure}
	\centering
	\includegraphics[width=8cm]{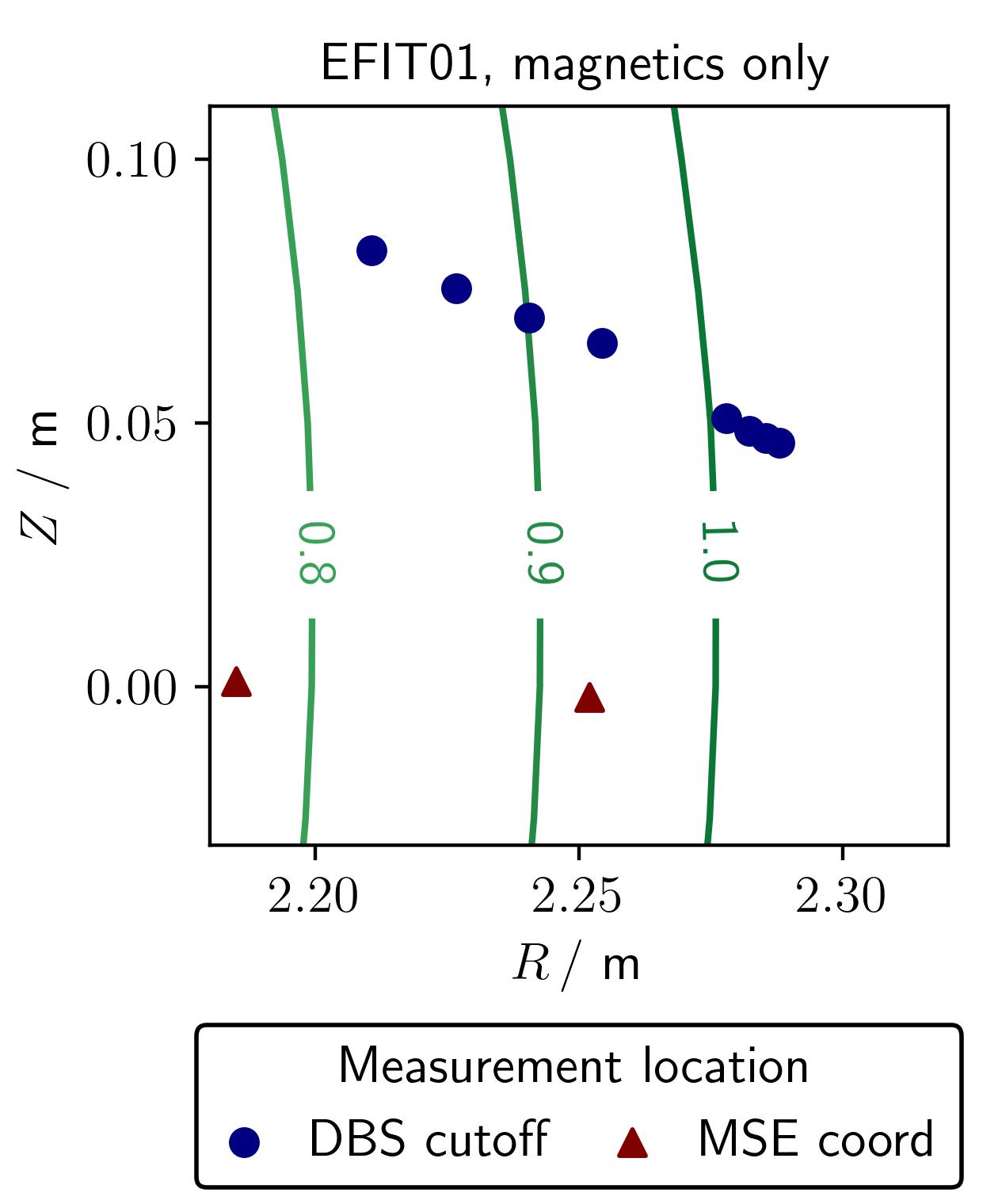}
	\caption{Locations of pitch angle measurements for MSE (triangles) and DBS (circles) for 1510 ms.}
	\label{fig:MeasurementLocation}
\end{figure}

\begin{figure*}
	\centering
	\includegraphics[width=15cm]{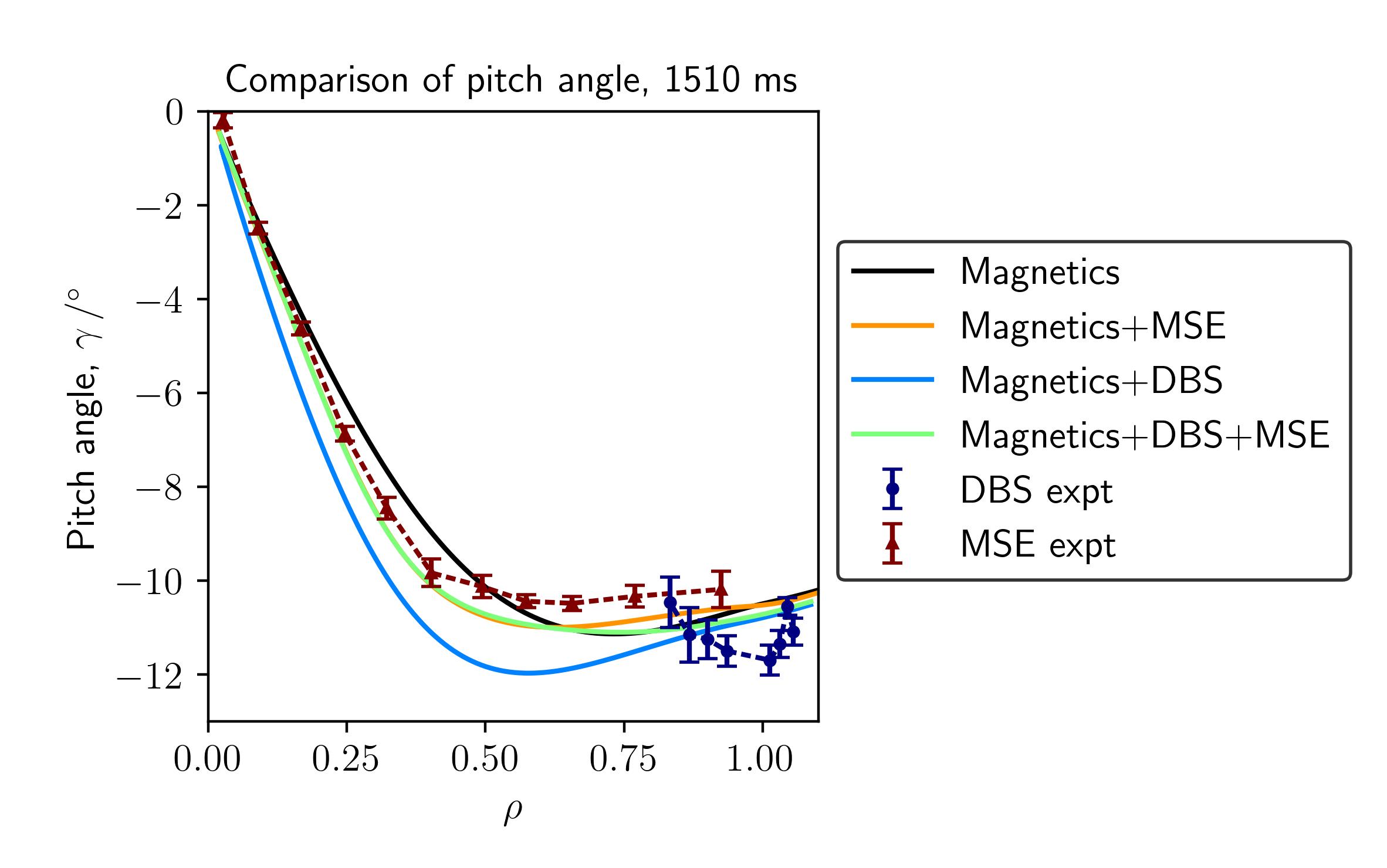}
	
	\caption{Motional-Stark effect (MSE) (dark red) and DBS (dark blue) measurements of the magnetic pitch angle, as compared with Equilibrium Fitting (EFIT) reconstructions using different constraints. We see that DBS measurements of the edge pitch angle are similar to that of MSE. As such, the DBS- and MSE-constrained EFIT reconstructions are also similar. Finally, we show that MSE and DBS can be used in conjunction to constrain the equilibrium reconstruction. Error bars in the DBS-measured pitch angle are from the uncertainty in the peak of the Gaussian fit, which is shown in Figure \ref{fig:gaussians}.}
	\label{fig:PitchComparison}
\end{figure*}

Since we assumed, in Section \ref{sec:BeamModel}, that the DBS signal comes entirely from the cutoff, determining the pitch angle scaling for a given DBS channel gives the actual pitch angle at the cutoff location, see Figure \ref{fig:MeasurementLocation}. 
In general, scattering along the beam path can be important. We simulated the line-integrated backscattered power, as a function of toroidal launch angle, with Scotty. For the shots studied in this paper, this line-integrated toroidal response is indistinguishable from that of cutoff-only mismatch attenuation. Hence, assuming the signal comes entirely from the cutoff is a good approximation. Since every frequency reaches cutoff at a different location, we now have localised measurements of pitch angle at different locations in the core, see Figure \ref{fig:PitchComparison}.
The edge pitch angle as measured by magnetics, DBS, and MSE are indeed similar, but with a slight difference between DBS and MSE. MSE measures a combination of the motional and radial electric fields \cite{Thorman:MSE:2018, Ko:MSE:2024}. Isolating the motional component is key for determining the pitch angle. However, the radial electric field can be large in the edge, making MSE measurements challenging. Furthermore, the radial resolution of MSE is poor at the edge, approximately on the magnitude of $\sim 10 \, {\rm cm}$, because it is determined by the intersection of the viewing volume with the neutral beam \cite{Levinton:MSE:1999,Rice:MSE:1997}. Hence, using DBS to measure the edge pitch angle might have advantages over MSE. Having measured the pitch angle with DBS, we then use it to constrain the equilibrium reconstruction, finding that the DBS-constrained equilibrium has a larger pitch angle than that from magnetics or EFIT. The constraints from DBS can be used in a standalone manner or in conjunction with MSE, potentially improving the fidelity of the reconstruction. Hence, we have established a proof of principle that DBS can indeed be used to measure the magnetic pitch angle.

\begin{figure*}
    \centering
    \includegraphics[width=13cm]{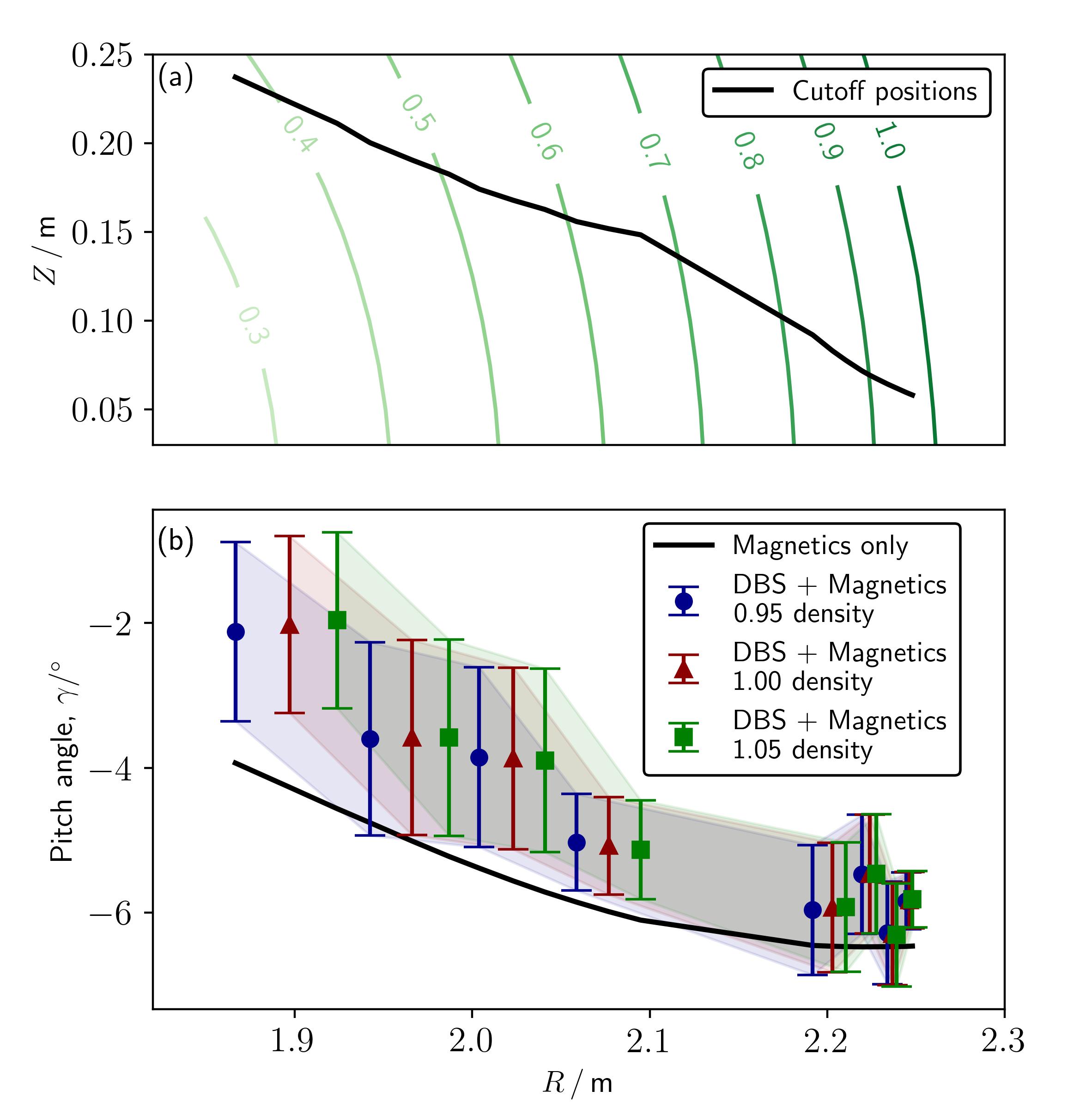}
    \vspace{-0.3cm}
    \caption{We now use DBS to infer the core pitch angle. To enable probing of the core, we use an equilibrium with a lower density, at 500 ms from the same set of shots. The cutoff locations of the eight DBS channels is shown in (a), together with contours of the flux surfaces, labelled by their normalised radial coordinates. In (b), we compare the pitch angles from the magnetics-only EFIT reconstruction with the DBS pitch-angle measurements. The latter consists of three sets of points, corresponding to using three different density profiles for our DBS analysis, to evaluate the effect of uncertainty in electron density on DBS measurements. Each set has eight points, corresponding to the eight DBS frequency channels. The uncertainty in electron density can arise from Thomson scattering measurements or from slight shot-to-shot variation. The red triangles correspond to using the electron density as measured by Thomson scattering, while the blue circles and green squares correspond to using a 5\% lower and 5\% higher electron density, respectively. The error bars are obtained from the errors propagated from the uncertainties in the Gaussian fits of experimental data. We find that the DBS-measured pitch angle is similar to that from magnetics near the edge; this is precisely where we expect the magnetics-only fit to be more accurate. The difference in pitch angle measured by DBS and magnetics is approximately a degree in the core, which cannot be accounted for with errors alone. Unfortunately, as MSE was not available at this time, we are unable to definitively establish DBS as measuring a more-accurate pitch angle in the core than magnetics. 
    }
	\label{fig:PitchAngleMeasurements500ms}
\end{figure*}

Having shown that DBS measurements of the magnetic pitch angle in the edge is consistent with that from magnetics and MSE, we now proceed to evaluate its applicability in the core. Since the shots studied had poor core access during flat top, see Figure \ref{fig:equilibrium500ms1510ms}, we study an earlier time during the set of shots, during ramp up. Equilibrium reconstruction during ramp up is more challenging as the current profile is still evolving. Moreover, we do not have MSE measurements during ramp up. Nonetheless, as a proof of principle, we find that the DBS is indeed able to yield measurements of pitch angle in the core, which differ from magnetics by one to two degrees, Figure \ref{fig:PitchAngleMeasurements500ms}. The implications of errors in the measurement of electron density will be discussed later in subsection \ref{subsec:ElectronDensityError}.

Our technique of using simultaneous toroidal DBS measurements to infer the local pitch angle is also applicable to the core of tokamak plasmas, which we discuss in a later section, see Figure \ref{fig:PitchAngleMeasurements500ms}. Regrettably, we do not have motional-Stark effect measurements of the pitch angle at the times when DBS had core access, which would have enabled cross comparison with a well-established technique. This comparison is beyond the scope of the current work. The following step would be to couple a beam-tracing code, like Scotty \cite{Hall-Chen:beam_model_DBS:2022}, with a Grad-Shfranov solver, such as FreeGSNKE \cite{Amorisco:FreeGSNKE:2024}, EFIT \cite{Lao:EFIT:1985, Appel:EFIT:2006}, or CHEASE \cite{Lutjens:CHEASE:1996}, such that the DBS measurements constrain the magnetic reconstruction at every iteration of the solver. That being said, having established DBS as a potential technique for measuring pitch angle, we proceed to show how to design a DBS system specifically for this purpose in the following section.

\section{Design considerations to improve the accuracy of a DBS pitch angle diagnostic}
\label{sec:DesignConsiderations}
Throughout this paper, we have used repeated shots to build the toroidal response one toroidal launch angle at a time. While this approach is well-supported by existing hardware, repeating shots is uneconomical and introduces errors due to shot-to-shot variation in plasma parameters. Designing a DBS system that is able to simultaneously measure multiple toroidal locations is thus key to measuring the magnetic pitch angle.

One possible approach is to set up a DBS system with in-shot toroidal steering, as was done for poloidal steering in previous work \cite{Chowdhury:DBS:2023}. By changing the toroidal launch angle via mirror steering quickly enough, such that the plasma equilibrium is effectively stationary during the sweep, the magnetic pitch angle can be determined. This would require a relatively modest upgrade of existing hardware. However, the drawback is that the temporal resolution of pitch-angle measurements would be compromised.

Another approach would be to install an array of DBS systems to enable simultaneous measurements at different toroidal launch angles. While this would constitute a more significant hardware investment, this would in-principle enable measurements of the pitch-angle with the high time resolution ($\lesssim 1$ms) of typical DBS and conventional reflectometry systems. Careful design might be required to avoid forward scattering from one DBS system to another as well as crosstalk between neighbouring antennas. We can, for example, carry out beam tracing for both antennas to ensure that the volumes of the receive and reciprocal emit beams do not overlap. To avoid crosstalk, the antennas can be placed arbitrarily far apart, which does not limit the technique proposed in this paper. If space constraints require the use of neighbouring antennas, a comprehensive cross-talk analysis is essential. This analysis should include, for example, near-field full-wave simulations and experimental measurements in vacuum. 

Nonetheless, a comprehensive study of the aforementioned hardware is beyond the scope of this paper. Throughout the rest of this section, we will examine some of the design considerations and put forward suggestions for best implementing such a system. In subsection \ref{subsec:accuracy}, we study how accurately the pitch angle can be measured. Next, we show how to optimise the effect of initial beam width and curvature on such measurements, to design a DBS system specially for measuring the magnetic pitch angle, subsection \ref{subsec:InitialBeamConditions}. In subsection \ref{subsec:OptimPolAngle}, we demonstrate that by changing the poloidal angle, we are able to control the range of optimal toroidal angles for the various frequencies. Finally, in subsection \ref{subsec:ElectronDensityError}, we discuss the impact of uncertainty in measurements of electron density and how to mitigate this.

\subsection{Accuracy of pitch-angle measurements using an antenna array}
\label{subsec:accuracy}
Pitch angle DBS relies on measuring the optimal toroidal launch angle to sub-degree accuracy, as seen in Figure \ref{fig:TorOptimNormComb}. This is especially important if we seek to have the same accuracy in measuring pitch angle as MSE. 

\begin{figure}[htbp]
    \centering
    \begin{subfigure}[b]{0.45\textwidth}
        \centering
        \begin{tikzpicture}[scale=1.0, every node/.style={font=\sffamily\small}]

                
            
            
            \def\optTor{3}
            \def\torWidth{1}
            \def\alphaErr{30}
           
            \def\torMax{\optTor+3*\torWidth}
            \def\torMin{\optTor-3*\torWidth}
        
            \def\axisHeight{5.5cm}
            \def\axisWidth{9.0cm}
            
            \begin{axis}[name=myaxis,
            width=\axisWidth, height=\axisHeight,
            at={(0.cm,0.cm)}, anchor=south west,
            xmin=\torMin-2, xmax=\torMax+2, ymin=-0.2, ymax=1.3,
            axis x line=middle, axis y line=middle,
            xtick=\empty, ytick=\empty,
            xlabel={$\varphi_{\rm t}$}, ylabel=$P$,
            xlabel style={at={(axis cs:\torMax-0.5,0.)},anchor=north},
            ylabel style={at={(axis cs:0.,1.3)},anchor=east},
            ]
             
                \begin{scope}[shift={(axis cs: -1.9, -0.1)}]
                
                    \draw[black, thick] (axis cs:5.0,1.0) -- (axis cs: 5.5,1.0);
                    \node[anchor=west] at (axis cs: 5.4,1.0) {ideal};
                    
                    \draw[black!\alphaErr, thick] (axis cs: 5.0,0.8) -- (axis cs: 5.5,0.8);
                    \node[anchor=west] at (axis cs: 5.4,0.8) {measured};
                    
                    \draw[rounded corners=0pt] (axis cs: 4.9,0.7) rectangle (axis cs: 7.7,1.1);
                
                \end{scope}
                
                \addplot [domain=-1:7, samples=200, thick, black] {
                1.0 * gaussian(\optTor,\torWidth)
                };
        
                \def\optTorErr{1.0}
        
                \addplot [domain=-1:7, samples=200, thick, black!\alphaErr] {
                1.0 * gaussian(\optTor+\optTorErr,\torWidth)
                };
                \draw[black!\alphaErr, dash pattern={on 7pt off 2pt on 1pt off 3pt}, thick] (axis cs:\optTor+\optTorErr,1.0) -- (axis cs:\optTor+\optTorErr,0.0);
                
                \draw[DarkGreen, <->, thick] (axis cs: \optTor, 0.3) -- (axis cs: \optTor+\optTorErr, 0.3) node[pos=0.8, below, yshift=0.0pt, DarkGreen]{$\Delta \varphi_{\rm t,opt}$};

                \draw[black] (axis cs:3.5,0.95) --++ (0,0.07) node[above,black]{ };

                \draw[black, dashed, thick] (axis cs:{\optTor-2*\torWidth},0) -- (axis cs:{\optTor-2*\torWidth},1.0);
                
                \draw[black, dashed, thick] (axis cs:{\optTor-2*\torWidth},0.135) -- (axis cs:0.,0.135)  node[left, black]{$\frac{P_{\max}}{\rme^2}$};

                \draw[black, dash pattern={on 7pt off 2pt on 1pt off 3pt}, thick] (axis cs:\optTor,1.0) -- (axis cs:\optTor,0.0)  node[below, black]{$\varphi_{\rm t,opt}$};

                \draw[black, dashed, thick] (axis cs:\optTor,1.0) -- (axis cs:0.,1.0)  node[left, black]{$P_{\max}$};
                
                \draw[black, decorate, decoration={brace, amplitude=5pt}, thick] (axis cs:{\optTor-2*\torWidth},1.05) -- (axis cs:\optTor,1.05) node[midway, above, yshift=1.5pt, black]{$\Delta \varphi_{\rm t,width} = 1$};
        
                \coordinate (midAnt) at (axis cs:\optTor,-0.5);
                \coordinate (focus) at (axis cs:\optTor,-1.4);
                
            \end{axis}
            
            
            \def\ratioarc{0.7}
            \def\radiusarc{\radius*\ratioarc}
            
            \def\errAngle{7.0}
        
            \pgfpointdiff{\pgfpointanchor{midAnt}{center}}
                     {\pgfpointanchor{focus}{center}}
            \pgfmathparse{veclen(\pgf@x,\pgf@y)}
            \edef\radius{\pgfmathresult pt}
            \def\numAnt{5.0}
            \def\spread{60.0} 
            \def\torStep{\spread/(\numAnt-1)}

            \coordinate (antErr) at ($(focus)+({90+2*(\torStep)+\errAngle}: \radius)$);
            \begin{scope}[shift={(antErr)}, rotate={2*(\torStep)+\errAngle}]
                \draw[black!\alphaErr, dashed, fill=OffWhite!\alphaErr] (-0.15,0) -- (0.15,0) -- (0.2,0.4) -- (-0.2,0.4) -- cycle;
            \end{scope}
            \draw[black!30, dashed] (antErr) -- (focus);
        
            \def\powAntList{2.9,3.7,2.6,1.25,1.4}

            \foreach \i [evaluate=\i as \currentPow using {{\powAntList}[\i]}] in {0,...,4} {
                \pgfmathsetmacro{\ang}{-\spread/2 + \i*(\spread/(\numAnt-1))}
                \coordinate (ant\i) at ($(focus) + (\ang+90: \radius)$);
                
                \node at ($(ant\i)+(\ang+90:0.4)+(0.0, \currentPow)$) [DarkRed]{$\pmb{\times}$};
                
                \begin{scope}[shift={(ant\i)}, rotate=\ang]
                    \draw[black, fill=OffWhite] (-0.15,0) -- (0.15,0) -- (0.2,0.4) -- (-0.2,0.4) -- cycle;
                \end{scope}
                \draw[black] (ant\i) -- (focus);
                
            }

            \fill[black] (focus) circle (1.5pt);
            
            \draw[DarkBlue, ->, thick] ($(focus)+({90-(\torStep)}: \radiusarc)$) arc ({90-(\torStep)}:90-2*\torStep:\radiusarc) node[pos=0.8, right, yshift=-5pt, DarkBlue]{$\Delta\varphi_{\rm t,step}$};
            
            \draw[DarkRed, ->, thick] ($(focus)+({90+2*(\torStep)}: \radiusarc)$) arc ({90+2*(\torStep)}:90+2*\torStep+\errAngle:\radiusarc) node[pos=0.8, left, yshift=-5pt, DarkRed]{$\Delta\varphi_{\rm t,err}$};

        \end{tikzpicture}
        \caption{DBS antenna array}
        \label{fig:PhasedArray}        

    \end{subfigure}
    \hspace{20pt}
    \begin{subfigure}[b]{0.45\textwidth}
        \centering
        \includegraphics[width=8cm]{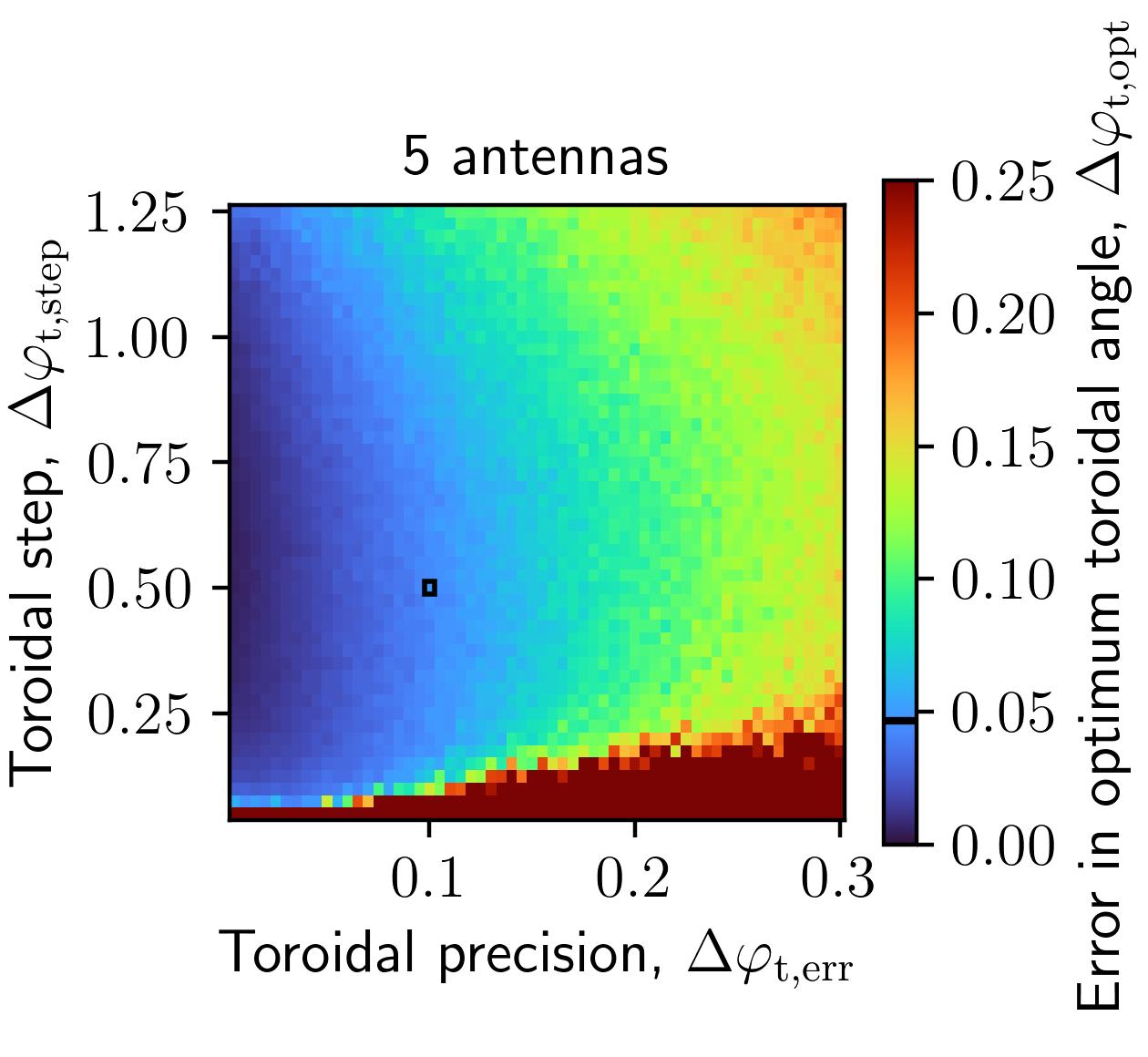}
        \caption{Monte Carlo simulations}
        \label{fig:MonteCarloGraph}
    \end{subfigure}
    \caption{We seek to determine the accuracy of DBS pitch-angle measurements, where our analysis is normalised to the width of the toroidal response, $\Delta \varphi_{\rm t,width} = 1$. (a) We use an array of five DBS antennas as an example. These antennas have the same poloidal launch angle but five different toroidal launch angles, $\varphi_{\rm t}$. The toroidal launch angles are intended to be equally spaced with a fixed toroidal step size, $\Delta \varphi_{\rm t,step}$. However, the toroidal steering has finite precision, which is taken to be the same for all antennas, $\Delta \varphi_{\rm t,err}$. We also assume a 3\% relative error in the backscattered power received by each antenna. If there were no errors, our measurements would yield the black line. However, with errors, a possible set of measurements is given by the red crosses. A Gaussian fit of these red crosses would give the grey line. The differences in the locations of the peaks of the black and grey lines is the error in the optimal toroidal launch angle, $\Delta \varphi_{\rm t, opt}$, exaggerated for illustration purposes. (b) This process of generating synthetic measurements and fitting them is repeated over 500 instances; the root-mean-square of the resulting ensemble defines the final estimated error $\Delta \varphi_{\rm t, opt}$. For the DBS system and plasma scenario at 1510 ms presented in this paper, the 67.5 GHz channel is indicated by the square marker in (b). In this case, $\Delta \varphi_{\rm t, err} \approx 0.3^\circ$, $\Delta \varphi_{\rm t, step} \approx 1.5^\circ$ and $\Delta \varphi_{\rm t,width}\approx 2.8^\circ$ (see Section \ref{sec:ExperimentalSetup}). The analysis predicts an error of $\Delta \varphi_{\rm t, opt} \approx 0.05 \approx 0.15^\circ$. With reference to Figure \ref{fig:TorOptimNormComb}(b), this translates to an error in the pitch angle of $\Delta \gamma \approx 4 \Delta \varphi_{\rm t, opt} \approx 0.56^\circ$. 
    }

    \label{fig:CombinedResults}
\end{figure}

We consider a system where there are simultaneous measurements at different toroidal launch angles, for example, an antenna array, see Figure \ref{fig:PhasedArray}. We consider five antennas with evenly spaced toroidal launch angles, with some error in these angles. We then carry out a Monte Carlo analysis with all quantities normalised to the $1/ \rme^2$ width of the toroidal response, which we refer to as the toroidal width, $\Delta \varphi_{t,width}$. The optimum range of toroidal launch angles, is such that they cover $\Delta \varphi_{t,width}$. This matches intuition; one should seek to spread the toroidal launch angles such that they cover a range of $\Delta \theta_{\rm m}$ in mismatch angles at the cutoff. In practice, we need an initial estimate of $\Delta \varphi_{t,width}$; which we obtain from Scotty simulations with the plasma equilibrium. 


We see that an array of five antennas is sufficient to get the accuracy of DBS-measured pitch angle to that of MSE, for the shots studied. Such a system could be implemented in existing tokamaks. 

\subsection{Optimising initial beam widths and curvatures} 
\label{subsec:InitialBeamConditions}
In the previous subsection, we saw that the best toroidal spread of measurements depends on the mismatch attenuation. Mismatch attenuation depends strongly on the probe beam's width and curvature at cutoff, which is in turn affected by the width and curvature at launch. Up to this point, we calculated these quantities from quasioptics of the particular DBS system we used \cite{Rhodes:DBS:2018}. While traditionally DBS designs should seek to be less sensitive to pitch angle matching \cite{Rhodes:DBS:2018, Hillesheim:DBS_MAST:2015, Hall-Chen:mismatch:2022} (larger $\Delta \theta_{\rm m}$), pitch-angle DBS benefits from being more sensitive to pitch angle matching (smaller $\Delta \theta_{\rm m}$). Since we seek to measure a small shift of the optimal toroidal launch angle, it is beneficial to have a smaller $\Delta \varphi_{\rm t,width}$ (corresponding to a small $\Delta \theta_{\rm m}$).

\begin{figure}
	\centering
	\includegraphics[width=8cm]{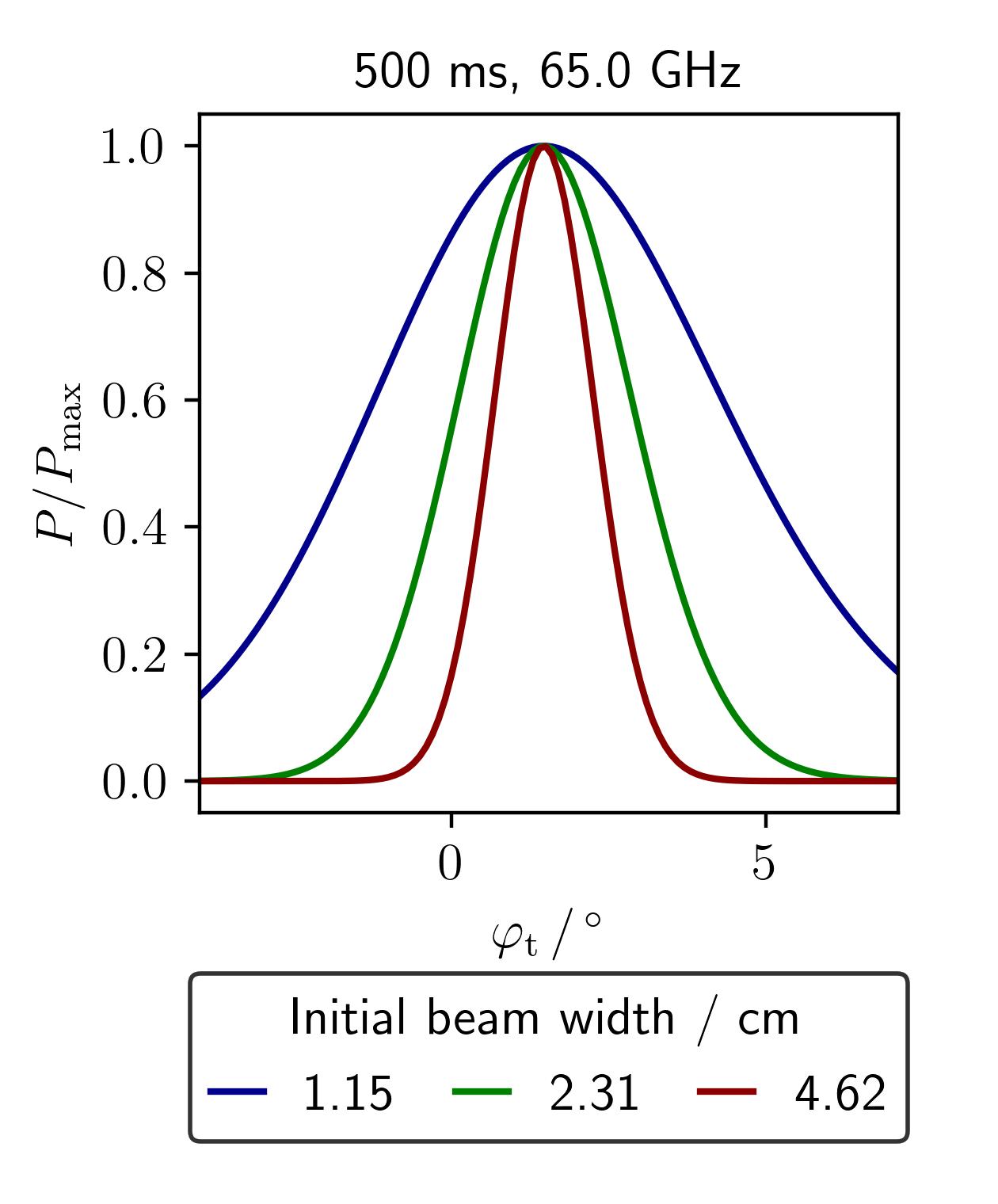}
	\caption{Simulated toroidal response $P/P_{\rm max}$ as a function toroidal launch angle $\varphi_{\rm t}$ for various beam widths. The curvature is fixed at -0.71 m$^{-1}$. By changing the initial beam conditions, we are able to control the width of the mismatch attenuation, and hence the toroidal width. Since pitch angle measurements depend on accurately determining the optimal toroidal launch angle (that which gives the maximum power), having a narrower toroidal width would give more accurate measurements.}
	\label{fig:PowerWidth65.0}
\end{figure}

We find that the mismatch tolerance can be reduced, in some cases by up to a factor of two over the existing DBS system, see Figure \ref{fig:PowerWidth65.0}. In this particular case, the simulated mismatch attenuation was reduced by keeping the launch curvature constant but increasing the launch width. If one were to assume vacuum propagation, it should be possible to arbitrarily decrease the mismatch tolerance by having a larger beam waist \cite{Hall-Chen:beam_model_DBS:2022, Hall-Chen:mismatch:2022}. However, we have found it difficult to get better than a factor of two reduction in the DIII-D plasma studied, no matter how we varied the width and curvature. Hence, we see that the plasma plays an important role in the evolution of the beam width and curvature, effectively imposing a lower limit on the size of mismatch attenuation.
Having a smaller mismatch attenuation reduces the required range of toroidally separated measurements, which would make toroidal steering via an oscillating mirror more achievable. However, in plasmas where the pitch angles, and thus the optimal toroidal launch angles, vary significantly over the different DBS frequencies, it might be possible for the mismatch attenuation to be so small that a given range of toroidal launch angles might not be able to simultaneously measure the pitch angles over all channels. Hence, proper optimisation with a beam-tracing code is imperative for designing such a system.

\subsection{Optimising the poloidal launch angle}
\label{subsec:OptimPolAngle}
\begin{figure*}
	\centering
	\includegraphics[width=15cm]{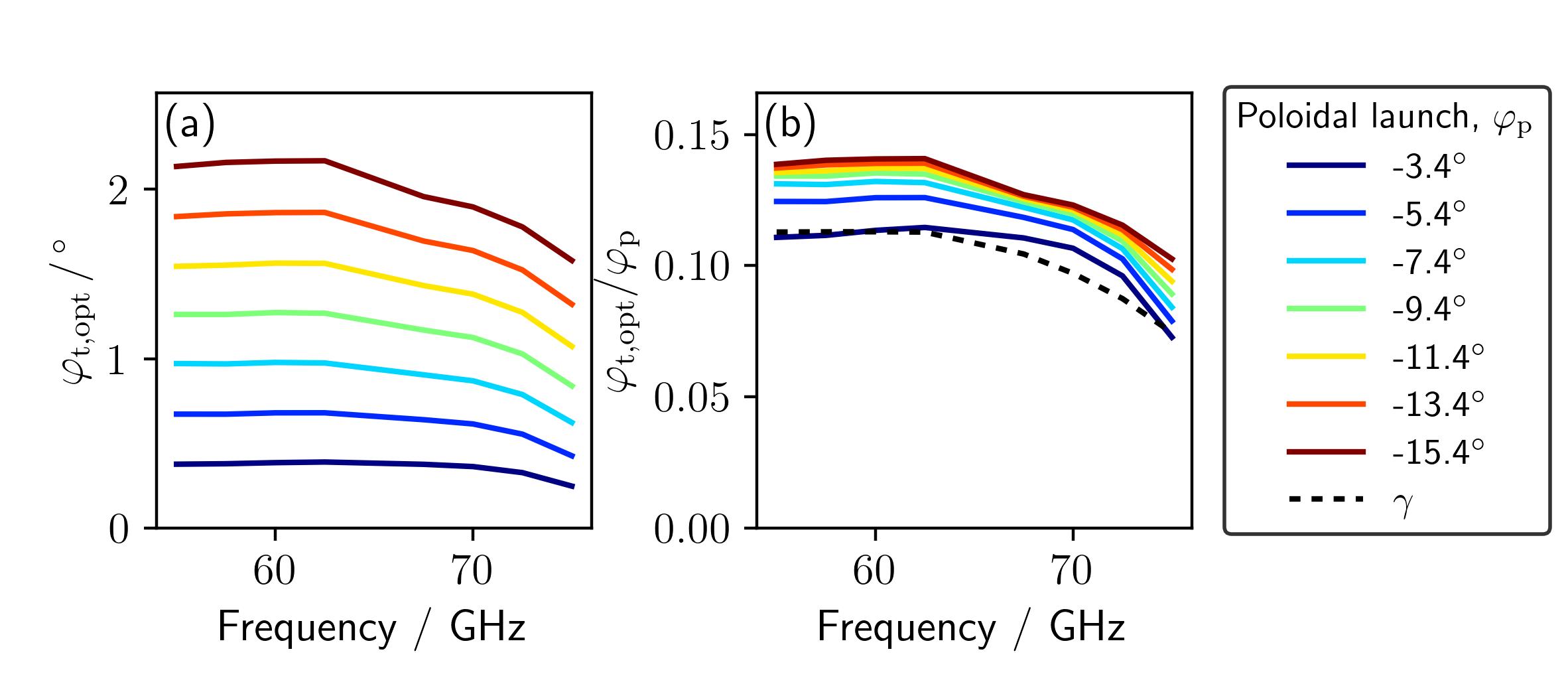}
	\caption{At 500ms, the optimal toroidal angle, $\varphi_{\rm t, opt}$, of all frequencies is larger when the poloidal launch, $\varphi_{\rm p}$, is larger (a). The ratio of $\varphi_{\rm t, opt}$ to $\varphi_{\rm p}$ (b) follows a qualitatively similar trend as the pitch angle at cutoff (dashed line). }
	\label{fig:TorOptimDiffPol}
\end{figure*}

All the experiments and simulations in the paper were carried out at fixed poloidal launch angle, $\varphi_{\rm p} = -11.4^\circ$. In this subsection, we investigate how changing this angle affects our measurements of pitch angle. As expected, we find that the optimal toroidal launch angles depend strongly on poloidal launch angle. Larger poloidal launch angles generally lead to larger optimal toroidal launch angles across all frequencies, see Figure \ref{fig:TorOptimDiffPol}.
As seen in subsection \ref{subsec:DbsMeasurementsPitchAngleScaling}, there is a linear dependence of optimal toroidal launch angle on pitch angle. Hence, if the optimal toroidal angle is larger, the same percentage difference in pitch angle will lead to a larger difference in optimal toroidal angle in degrees, which would be easier to measure. Similarly, larger poloidal angles lead to higher wavenumbers at cutoff for a given frequency; mismatch attenuation is stronger, that is $\Delta \theta_{\rm m}$ is smaller, at larger wavenumbers \cite{Hall-Chen:mismatch:2022}. With these consideration, larger poloidal launch angles should be favoured. However, higher wavenumbers result in lower backscattered signals, which may make detection difficult. At larger poloidal angles, the localisation of scattered power to the cutoff is also weaker, which may lead to worse spatial resolution. Furthermore, using an oscillating mirror would favour smaller changes in optimal toroidal launch angle, which would enable the mirror to be oscillated at higher frequencies, increasing the time resolution of pitch angle measurements. Designing a pitch angle DBS would have to balance these trade-offs and select an intermediate poloidal launch angle.

\subsection{Error introduced by electron density measurements}
\label{subsec:ElectronDensityError}

An important source of error is the uncertainty in the electron density measurements. Generally speaking, if the electron density is higher, the cutoff position will be further out, see Figure \ref{fig:PitchAngleMeasurements500ms}. In this set of shots, there are two sources of error. The first arises from the error of Thomson scattering measurements. The second, is because of the slight differences in the operating conditions of the repeated shots, that result in physically different electron densities. If the DBS array approach is adopted, we can effectively eliminate the second source of error due to the repeating of shots. 

We expect such error to be especially important when probing the core, as they would compound when doing beam tracing. Moreover, the core density gradient is likely less steep than that at the edge; for the same percentage error in electron density, we expect a large error in cutoff location for core measurements than edge measurements. We repeated our simulations for the 500 ms case with 5\% higher and lower densities, scaled globally, see Figure \ref{fig:PitchAngleMeasurements500ms}; and found that it did not significantly impact DBS measurements of pitch angles at the edge. However, in the core, these errors can affect the expected measurement locations by centimeters, but the measured value of the pitch angle did not change significantly. Consequently, to improve spatial accuracy, X-mode operation is preferable to O-mode. This is because of the X-mode cutoff frequencies have a stronger dependence on radial position than O-mode, see Figure \ref{fig:equilibrium500ms1510ms}. This increased gradient results in the cutoff position of X-mode being better resolved in the radial direction. 

The spatial resolution of a pitch-angle DBS is dependent on accurate measurements of density and will likely need to be supplemented by other systems, such as profile reflectometry and interferometry, in future fusion energy systems. 

\section{Conclusion}
\label{sec:conclusion}
In this paper, we proposed a new technique for measuring the magnetic pitch angle using DBS. We applied this technique to a series of eleven repeated DIII-D discharges where the DBS toroidal launch angle was systematically varied between shots, enabling us to experimentally determine the optimal toroidal launch angle. This optimal toroidal angle is that where the backscattered power is maximised because the beam wavevector is perpendicular to the magnetic field at cutoff. We compared the experimentally calculated optimal toroidal launch angle with Scotty simulations where the pitch angle was computationally scaled. Matching experiment and simulations, together with assuming that the DBS signal was well-localised to the cutoff location, we were then able to calculate the local magnetic pitch angle to an uncertainty of $\approx 0.5 ^\circ$. Since MSE measurements of pitch angle were only available at one particular time, we compared that to DBS measurements of the same, finding reasonable agreement, of within a degree, across all DBS channels ($\rho \simeq 0.8$--$1.0$).

While our comparison with MSE was at the edge, DBS can potentially measure the magnetic pitch angle in the core. During ramp up, when the DBS beams penetrated further into the core, we found that DBS measurements of pitch angle differed from magnetics-only reconstruction by 1--2 degrees, with the caveat that equilibrium reconstruction during ramp up is challenging and requires work beyond the scope of this paper. As a result of shallower density gradients in the core and longer paths of the DBS probe beams, the accuracy of the density profile is especially important for finding the location of DBS measurements. However, small changes in the density profile are less important for the value of the pitch angle measured. 

We studied the effect of noise in the backscattered power and the uncertainty in toroidal launch angle on the error in pitch angle measurements. For existing hardware on DIII-D, we found that the latter was much more significant. To further reduce uncertainty, we suggest developing a specialised pitch-angle DBS capable of simultaneous toroidally separated measurements, rather than varying the toroidal angle from shot to shot, as shots are not perfectly repeatable and repetition can be cumbersome. Such a specialised pitch-angle DBS can be further optimised by selecting appropriate poloidal launch angles and Gaussian beam properties, which are complicated and strongly dependent on plasma properties. The considerations for selecting the initial beam width and curvature are different, sometimes entirely opposite, from that needed for traditional DBS systems for measuring density fluctuations and flows. DBS measurements of pitch angle, especially when designing systems for new machines, thus necessitates advanced modelling of beam properties, which are now easily and quickly available with Scotty \cite{Hall-Chen:beam_model_DBS:2022}. 

Using DBS as a pitch-angle diagnostic has several advantages. It is non-perturbative, does not require neutral beams, measures the pitch angle rather than its time derivative, has a high temporal resolution, is spatially localised, and being a microwave technique, has front-end hardware that is robust to damage from neutron irradiation. Hence, we expect this technique to extrapolate well to burning plasmas of future fusion energy systems, potentially providing critical pitch-angle measurements for control.

\ack {This work was partially funded by the FEAT SRTT, A*STAR. AK Yeoh was funded by a National Science Scholarship from A*STAR, Singapore. This material is also based upon work supported by the U.S. Department of Energy, Office of Science, Office of Fusion Energy Sciences, using the DIII-D National Fusion Facility, a DOE Office of Science user facility, under Awards DEFC02-04ER54698, DE-SC0019352, DE-AC02-09CH11466, and DE-AC52-07NA27344. This work was also in part supported by a grant from the Engineering and Physical Sciences Research Council (EPSRC) [EP/R034737/1]. J. Ruiz Ruiz has been supported by EPSRC [EP/W026341/1]. We thank V.N. Duarte for helpful discussions.}\\

\noindent\textbf{Disclaimer}. This report was prepared as an account of work sponsored by an agency of the United States Government. Neither the United States Government nor any agency thereof, nor any of their employees, makes any warranty, express or implied, or assumes any legal liability or responsibility for the accuracy, completeness, or usefulness of any information, apparatus, product, or process disclosed, or represents that its use would not infringe privately owned rights. Reference herein to any specific commercial product, process, or service by trade name, trademark, manufacturer, or otherwise does not necessarily constitute or imply its endorsement, recommendation, or favoring by the United States Government or any agency thereof. The views and opinions of authors expressed herein do not necessarily state or reflect those of the United States Government or any agency thereof. The United States Government retains a non-exclusive, paid-up, irrevocable, world-wide license to publish or reproduce the published form of this manuscript, or allow others to do so, for United States Government purposes.

\appendix

\section{Effect of wavenumber spectrum on toroidal response}
\label{sec:wavenumberSpectrum}

\begin{figure*}[h!]
	\centering
	\includegraphics[width=8cm]{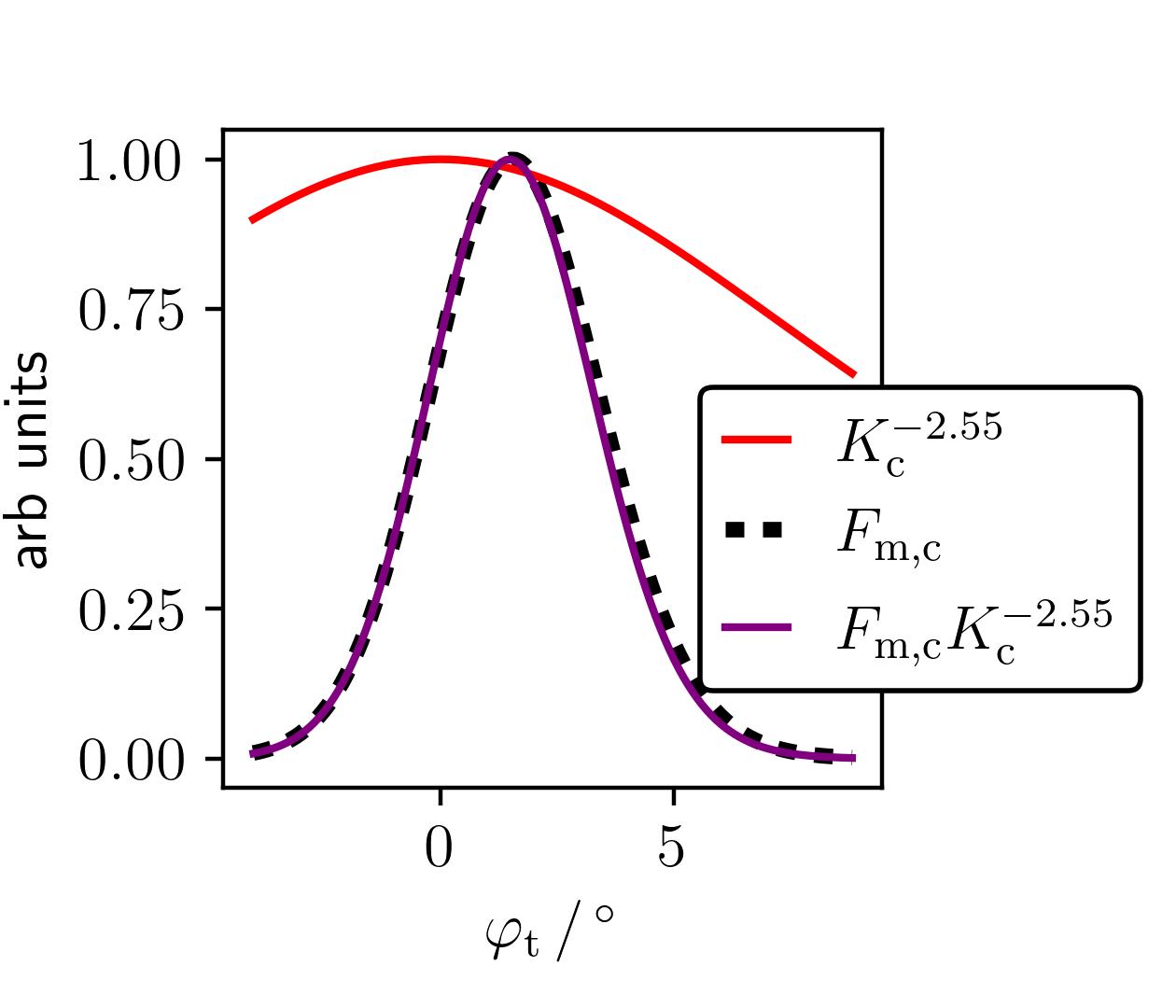}
	\caption{Various contributions to the backscattered power, see equation (\ref{eq:PowerReceived}), at the cutoff location for 55 GHz channel at time 500 ms. 55 GHz corresponds to the case where the wavenumber changes the most relatively when the toroidal launch angle $\varphi_{\rm t}$ changes. The mismatch term $F_{\rm m,c}$ given in equation (\ref{eq:FilterMismatch}). The combined influence from the turbulence spectrum, is estimated as $K_{\rm c}^{-2.55}$, see Figure 6 of \cite{Pratt:spectrum:2023}. The changing wavenumber $K_{\rm c}^{-2.55}$ has a minimal influence on the resulting plot, $F_{\rm m,c} K_{\rm c}^{-2.55}$.  All plots here are normalized to their maximum values.}
	\label{fig:wavenumberInfluence}
\end{figure*}

The turbulence spectrum affects the backscattered power through equation (\ref{eq:PowerReceived}); here, we justify our choice of excluding the turbulence spectrum in our analysis. The turbulence spectrum can be approximated by a power-law dependence on wavenumber \cite{Schekochihin:spectrum:2008, Schekochihin:spectrum:2009, Barnes:spectrum:2011, Pratt:spectrum:2023}. After running multiple simulations for the shots studied, we find that the turbulence spectrum has no significant effect on the toroidal response. We illustrate this using the 55 GHz channel at 500 ms, which has the largest relative change in wavenumber at cutoff, see Figure \ref{fig:CutoffWavenumberVSToroidalScaling}. We estimate the critical exponent to be $k_{\perp}^{-2.55}$, as seen from Figure 6 of previous work \cite{Pratt:spectrum:2023}. Using the Bragg condition $\mathbf{k}_{\rm \perp}=-2\mathbf{K}$, the effect of the turbulence spectrum on measurements can be written as a function of probe beam's wavenumber at cutoff, $K_{\rm c}^{-2.55}$. The effect of the turbulence spectrum on the toroidal response is indeed negligible, see Figure \ref{fig:wavenumberInfluence}. Nonetheless, this might not be the case for other plasma scenarios and tokamaks. If the effect of the spectrum is significant, the toroidal response might not be Gaussian, and other fitting techniques might need to be explored.
 
\section*{References}
\bibliographystyle{iopart-num}
\bibliography{references}
\end{document}